\documentclass{aa}

\usepackage{natbib}
\usepackage{graphicx}
\usepackage{amssymb}

\begin{document}

\newcommand{\3}{\ss}
\newcommand{\n}{\noindent}
\newcommand{\eps}{\varepsilon}
\newcommand{\be}{\begin{equation}}
\newcommand{\ee}{\end{equation}}
\newcommand{\bl}[1]{\mbox{\boldmath$ #1 $}}
\def\ba{\begin{eqnarray}}
\def\ea{\end{eqnarray}}
\def\de{\partial}
\def\msun{M_\odot}
\def\div{\nabla\cdot}
\def\grad{\nabla}
\def\rot{\nabla\times}
\def\ltsima{$\; \buildrel < \over \sim \;$}
\def\simlt{\lower.5ex\hbox{\ltsima}}
\def\gtsima{$\; \buildrel > \over \sim \;$}
\def\simgt{\lower.5ex\hbox{\gtsima}}

\title{Formation of giant planets and brown dwarfs on wide orbits}
\titlerunning{Formation of giant planets and brown dwarfs}

\author{Eduard~I.~Vorobyov, \inst{1,2} }
\authorrunning{E.~I.~Vorobyov}

\offprints{E.~I.~Vorobyov} 
   
\institute{University of Vienna, Institute of Astrophysics,  Vienna, 1180, Austria; 
\email{eduard.vorobiev@univie.ac.at}
\and
Research Institute of Physics, Southern Federal University, Stachki Ave. 194, Rostov-on-Don, 
344090 Russia }

\date{}

\abstract
{We studied numerically the formation of giant planet (GP) and brown dwarf (BD) embryos 
in gravitationally unstable protostellar disks and compared our findings with directly-imaged, 
wide-orbit ($\ga 50$~AU) companions known to-date. }
{The viability of the disk fragmentation scenario for the formation
of wide-orbit companions in protostellar disks around (sub-)solar mass stars was investigated.
The particular emphasis was paid to the survivability of GP/BD embryos formed via disk 
gravitational fragmentation.}
{We used numerical hydrodynamics simulations of disk formation and evolution
with an accurate treatment of disk thermodynamics. The use of the thin-disk limit allowed us 
to probe the long-term evolution of protostellar disks, starting from the gravitational
collapse of a pre-stellar core and ending in the T Tauri phase after at least 1.0~Myr of
disk evolution.
We focused on models that produced wide-orbit GP/BD embryos, which opened a gap in the disk 
and showed radial migration timescales similar to or longer than the typical disk lifetime.}
{While disk fragmentation was seen in the majority of our models, only 6 models out of 60 
revealed the formation of quasi-stable, wide-orbit GP/BD embryos. 
The low probability for the fragment survival is caused by efficient
inward migration/ejection/dispersal mechanisms which operate in the embedded phase of
star formation. We found that only massive and extended protostellar disks 
($\ga 0.2~M_\odot$),  experiencing gravitational 
fragmentation not only in the embedded but also in the T~Tauri 
phases of star formation, can form wide-orbit companions. 
Disk fragmentation produced GP/BD embryos with masses in the 3.5--43~$M_{\rm J}$ 
range, covering the whole mass spectrum of directly-imaged, wide-orbit companions to
(sub-)solar mass stars. 
On the other hand, our modelling failed to produce embryos on orbital distances $\la 170$~AU, 
whereas several directly-imaged companions were found at smaller orbits down to a few~AU.
Disk fragmentation also failed to produce wide-orbit companions around stars with 
mass $\la 0.7~M_\odot$, in disagreement with observations.} 
{Disk fragmentation is unlikely to explain the whole observed spectrum of 
wide-orbit companions to (sub-)solar-mass stars and other formation mechanisms, e.g., 
dynamical scattering of closely-packed companions onto wide orbits, 
should be invoked to account for companions at orbital distance from a few tens to 
$\approx 150$~AU and wide-orbit companions with masses of the host star $\le 0.7~M_\odot$. 
Definite measurements of 
orbit eccentricities and a wider sample of numerical models are needed to distinguish 
between the formation scenarios of GP/BD on wide orbits.}

\maketitle

\section{Introduction}

With the detection of giant planets (GPs) and brown dwarfs (BDs) on orbital distances of the order
of tens to hundreds AU \citep[e.g.][]{Marois08,Kalas08,LJK10,Schmidt08}, gravitational 
instability and fragmentation of protostellar disks has gained a renewed interest as one of the 
likely mechanisms that can explain the formation of wide-orbit companions 
\citep[e.g.][]{Boss03,Boss11,Stamatellos09,Boley09,DR09,Kratter10a,Kratter10b,VB2010b}.
Measurements of disk masses in the T~Tauri phase of stellar evolution
yielded a few candidates with disks masses as massive as 0.1--0.5~$M_\odot$ 
\citep[][]{Eisner08,Andrews09,Isella09}.
Such massive disks are expected to be more frequent 
in the earlier, embedded phase of star formation 
\citep[e.g.][]{Eisner05,Jorgensen09,Vor2011,Eisner12},
reinforcing the viability of disk fragmentation
at least in this early evolutionary stage.

Numerical hydrodynamics simulations and semi-analytical studies  seem to converge on
that disk fragmentation is feasible at distances greater than a few tens of AU from the 
central star \citep{Mayer07,Boley09,Rice10,VB2010a,Zhu2012}, where the 
cooling time becomes shorter than the local dynamical
timescale \citep{Gammie01, Johnson03,Rice03}. Although the exact distance beyond which
disk fragmentation may operate is still under debate \citep{Rafikov05, Nero09, Meru2011, 
Meru2012}, it now becomes evident that parental cores must have enough mass and angular momentum
to form extended disks with mass sufficient to drive the local 
Toomre $Q$-parameter below unity \citep[e.g.][]{VB2010a,Vor2011}.

It has recently become evident that the formation of massive fragments via disk gravitational fragmentation
does not guarantee that the fragments will ultimately evolve into GPs/BDs on wide orbits.
Gravitational instability in the embedded phase of star formation is strong, 
fuelled with a continuing infall of gas from a parent cloud core, and resultant gravitational
and tidal torques from spiral arms are rampant.  As a consequence, the majority of the
fragments are torqued into the inner disk regions, probably producing a population of close-in
terrestrial and icy giants due to tidal disruption/tidal downsizing  
\citep{Nayakshin10,Boley10,Cha11}, or driven directly onto the star,
triggering intense accretion and luminosity bursts similar in magnitude to FU Orionis eruptions
\citep{VB06,VB2010a}. Some of the fragments are dispersed by tidal torques exerted by the spiral
arms before they can dissociate molecular hydrogen in their interiors and contract to planetary-sized
objects \citep{Boley10,Vor2011a,Nayakshin11}. A few fragments may be scattered away from the disk
via many-body gravitational interaction with outer fragments or fully formed sub-stellar objects,
 producing freely floating sub-stellar objects \citep{Stamatellos09,BV12}.

Therefore, the question of whether fragments can settle into stable 
orbits at distances where the form, from several tens to hundreds AU, remains to be open. 
\citet{Baruteau11} studied
planet migration in graviturbulent disks with mass $0.4~M_\odot$ 
and argued that Jupiter-mass planets (or higher) initially placed at 100~AU 
migrate inward on timescales of $10^4$~yr and are unlikely to stay on wide orbits. 
\citet{Michael11} found even faster 
migration timescales of $10^3$~yr for a Jupiter-mass planet in a $0.14~M_\odot$ disk,
though the planet may stall near the inner Lindblad resonance of the dominant spiral mode.
Other numerical hydrodynamics simulations \citep{VB06,VB2010a,Cha11,Machida2011} also
revealed fast inward migration of the forming fragments in the disk.
On the other hand, \citet{VB2010b} studied the long-term 
($\sim$~several Myr) evolution of fragmenting protostellar
disks in the thin-disk limit and found that while most disks indeed fail to produce 
stable companions on wide orbits, in agreement with previous studies,
a small subset of models can form GPs at distances of the order of tens to
hundreds AU. They concluded that only those fragments that happen to form in the late embedded phase,
when gravitational instability and associated torques are getting
weaker, may open a gap in the disk and mature into GPs on wide orbits.
The low probability for survival of the fragments formed via disk gravitational fragmentation 
was also confirmed by \citet{Zhu2012}, who studied numerically 
two-dimensional disks subject to mass loading and found that only 3 fragments out of 13 became massive
enough to open a gap in the disk and essentially stopped migrating. 

In this paper, we improve the model of \citet{VB2010b} by including a detailed thermal balance in 
the protostellar disk, thus removing the barotropic relation closure adopted in \citet{VB2010b} which
is known to overpredict the number of fragments \citep[e.g.][]{Bate09,Stamatellos09b}.
In our numerical hydrodynamics simulations, we form disks self-consistently 
during the gravitational collapse of pre-stellar cores and not introduce them artificially. This
allows us to determine the range of masses and angular momenta in prestellar cores 
for which disk fragmentation and formation of wide-orbit companions can take place. 
We also avoid replacing
fragments with point sink particles, thus studying the evolution of GP/BD embryos
rather than fully formed planetary- or sub-stellar objects. This is done to avoid a premature 
introduction of sink particles, the effect that can influence the number of surviving 
fragments due to essential indestructibility of point-sized objects. 
On the other hand, the formation of planetary/sub-stellar-sized objects 
is not resolved in the current approach, which may affect the short-term survivability of the fragments.
We focus on models that produce planetary- or sub-stellar-mass embryos on {\it quasi-stable} 
orbits at radial distances where disk fragmentation takes place ($\ga 50$~AU).

The organization of this paper is as follows. A brief description
of the numerical model is provided in Section~\ref{model}. 
The parameter-space study of disk fragmentation and fragment survival is
provided in Section~\ref{fragmentation}. Formation of GP and BD
embryos in wide orbits is presented in Sections~\ref{planet} and \ref{browndwarf}, respectively.
Feasibility for the formation of multiple companions in wide orbits
is discussed in Section~\ref{multiple}. Comparison of numerical results 
with observed sub-stellar objects
in wide orbits is performed in Section~\ref{comparison}. The main results
are summarized in Section~\ref{summary}.

\section{Model description}
\label{model}
Our numerical model is described in detail in 
\citet{VB2010b} and is briefly reviewed below for the reader's convenience.
We use numerical hydrodynamics simulations in the thin-disk approximation to 
compute the gravitational collapse of pre-stellar cores of various initial mass
and angular momentum. This approximation
is an excellent means to calculate the evolution for many
orbital periods and many model parameters, and its justification
is provided in \citep{VB2010a}. To avoid too small time steps, 
we introduce a ``sink cell'' at $r_{\rm sc}=6$~AU and 
impose a free boundary condition such that the matter is allowed to flow out of 
the computational domain but is prevented from flowing in. 
The sink cell is dynamically inactive; it contributes only to the total gravitational 
potential and secures a smooth behaviour of the gravity force down to the stellar surface.
We monitor the gas surface density in the sink cell and 
when its value exceeds a critical one for the transition from 
isothermal to adiabatic evolution, we introduce a central point-mass object.

The simulations continue into the embedded phase of star formation, during which
a protostellar disk is formed. In this stage, the disk is subject to intense mass loading from 
the remnant of the initial pre-stellar core---the so-called envelope. The self-consistent
disk-envelope interaction is a key feature of our model allowing us to observe repetitive episodes
of disk fragmentation in some models. During the diks evolution, 
90\% of the gas that crosses the inner boundary 
is assumed to land onto the central object plus the sink cell. 
The other 10\% of the accreted gas is assumed to be carried away with protostellar jets.
The simulations are terminated in the late T Tauri phase 
after more than one Myr of disk evolution 
when nearly all of the envelope material has accreted onto the resulting star plus disk system.

Models presented in this paper are run on a polar coordinate
($r, \phi$) grid with $512\times512$ zones. The radial points are logarithmically spaced,
with the innermost cell outside the central sink having size
0.07--0.1~AU  depending on the cloud core size (i.e., the radius of
the computational region). The latter varies in the 0.025--0.12~pc 
(5000--24,000 AU) limits. The radial and azimuthal resolution
are $\la 1.0$ AU at a radial distance $r\la 100$~AU. 

The Truelove criterion states that the local Jeans length must be resolved by at 
least four numerical cells in order to correctly capture disk fragmentation \citep{Truelove99}. 
For a circular fragment of radius $R_0$ with surface density inversely 
proportional to radius,
$\Sigma=\Sigma_0 R_0/r$, the kinetic energy due to random motions  can be expressed as
\begin{equation}
E_{\rm kin} = \pi R_0^2 \Sigma_0 \langle v^2 \rangle,
\end{equation}
where $\Sigma_0$ is the surface density at the fragment-disk interface. 
The velocity dispersion of  a thin disk with two translational degrees
of freedom\footnote{Our disks are not razor thin, but are characterized by a vertical scale height 
increasing outward, as occurs in flared disks. Taking into account the vertical degree of freedom 
changes the corresponding Jeans length by a factor of only 1.5.} is
$\langle v^2 \rangle = 2{\cal R} T_0/\mu$, where  $\cal R$ is the universal gas
constant, $\mu=2.33$ is the mean molecular weight, and $T_0$ is the gas midplane temperature 
in the fragment. 

The corresponding gravitational energy of the fragment is 
\begin{equation}
E_{\rm gr}= - 2 \pi \int \limits_0 \limits^R r g_{\rm r} \Sigma r dr = - 2 \pi^2 G \Sigma_0^2 R_0^3,
\end{equation}
where we have taken into account that $g_{\rm r}=\pi G \Sigma$ for a disk with 
$\Sigma \propto r^{-1}$ \citep[see][p.77]{BT87}. 
The resulting Jeans length $R_{\rm J}$ is 
calculated from the virial theorem $2E_{\rm kin} + E_{\rm g}=0$ as
\begin{equation}
R_{\rm J} = {\langle v^2 \rangle \over \pi G \Sigma_0}.
\label{RJeans}
\end{equation}
Fragments usually condense out of densest sections of spiral arms. The typical 
surface densities and temperatures in spiral arms do not exceed 100~g~cm$^{-2}$ and 
100~K (see Figures~\ref{fig2}, \ref{fig6}, \ref{fig8} in this paper and \citet{Vor2011}).
Adopting these values for $\Sigma_0$ and $T_0$, the corresponding Jeans length 
is $R_{\rm J}\approx 20$~AU.  

In models showing disk fragmentation, the radial and azimuthal grid resolution
at $r=100$~AU is $\approx 1.0$~AU and
the Jeans length is resolved by roughly 20 grid zones in each coordinate direction.
On our logarithmically spaced radial grid,
the Truelove criterion is expected to break only at $r\ga 500$~AU where
the grid resolution starts to exceed 5.0~AU. 
Fragmentation takes place mostly at radial distances from a few tens to a few hundreds AU. 
Fragments that are seen in our models at larger distances are most likely scattered from the 
inner disk regions due to gravitational interaction with other fragments\footnote{In the most
extreme cases, some of the fragments may even be ejected from the disk into 
the intracluster medium \citep{BV12}.}. 
The radii of the survived fragments (see the ninth column in Table~\ref{table2}) lie between
10 and 20 AU, implying that the fragments are resolved on the two-dimensional mesh 
by {\it at least} 30--60 grid zones in the inner 500~AU. 
We therefore conclude that the numerical resolution
in our models is sufficient to capture disk fragmentation correctly.
On the other hand, contraction of the survived fragments to planetary-sized objects
cannot be modelled in the current approach. This may have consequences for the survivability of the
fragments, resulting in an increased probability of tidal destruction of AU-sized objects as compared
to planetary-sized ones. Accreting sink particles 
are needed to correctly follow the evolution of fully formed giant planets and brown dwarfs.

\subsection{Basic equations}
In the paper of \citet{VB2010a}, a barotropic equation of state was used to close 
the equations of hydrodynamics. In this work, we include detailed thermal physics
in our model, the main concepts of which are briefly reviewed below.
The basic equations of mass, momentum, and energy transport  are
\begin{equation}
\label{cont}
\frac{{\partial \Sigma }}{{\partial t}} =  - \nabla_p  \cdot 
\left( \Sigma \bl{v}_p \right),  
\end{equation}
\begin{eqnarray}
\label{mom}
\frac{\partial}{\partial t} \left( \Sigma \bl{v}_p \right) &+& \left[ \nabla \cdot \left( \Sigma \bl{v_p}
\otimes \bl{v}_p \right) \right]_p =   - \nabla_p {\cal P}  + \Sigma \, \bl{g}_p + \\ \nonumber
& + & (\nabla \cdot \mathbf{\Pi})_p, 
\label{energ}
\end{eqnarray}
\begin{equation}
\frac{\partial e}{\partial t} +\nabla_p \cdot \left( e \bl{v}_p \right) = -{\cal P} 
(\nabla_p \cdot \bl{v}_{p}) -\Lambda +\Gamma + 
\left(\nabla \bl{v}\right)_{pp^\prime}:\Pi_{pp^\prime}, 
\end{equation}
where subscripts $p$ and $p^\prime$ refers to the planar components $(r,\phi)$ 
in polar coordinates, $\Sigma$ is the mass surface density, $e$ is the internal energy per 
surface area, 
${\cal P}$ is the vertically integrated gas pressure calculated via the ideal equation of state 
as ${\cal P}=(\gamma-1) e$ with $\gamma=7/5$, 
$Z$ is the radially and azimuthally varying vertical scale height
determined in each computational cell using an assumption of local hydrostatic equilibrium,
$\bl{v}_{p}=v_r \hat{\bl r}+ v_\phi \hat{\bl \phi}$ is the velocity in the
disk plane, $\bl{g}_{p}=g_r \hat{\bl r} +g_\phi \hat{\bl \phi}$ is the gravitational acceleration 
in the disk plane, and $\nabla_p=\hat{\bl r} \partial / \partial r + \hat{\bl \phi} r^{-1} 
\partial / \partial \phi $ is the gradient along the planar coordinates of the disk. 
We note that the adopted value of $\gamma$ neglects a possible stiffening of the equation of state
at low temperatures ($<100$~K) where the ratio of specific heats may approach a value typical
for a monatomic gas, $\gamma=5/3$ \citep{MI2000}. The adopted equation of state may be important
for disk gravitational instability \citep{Boley07}, though the recent study of \citet{Zhu2012} found
that fragmentation of two-dimensional disks does not depend sensitively on $\gamma$ in the range from
7/5 to 5/3.

Turbulent viscosity is taken into account via the viscous stress tensor 
$\mathbf{\Pi}$  expressed as
\begin{equation}
\mathbf{\Pi}=2 \Sigma\, \nu \left( \nabla \bl{v} - {1 \over 3} (\nabla \cdot \bl{v}) \mathbf{e} \right),
\label{stressT}
\end{equation}
where $\mathbf{e}$ is the unit tensor and $\nabla \bl{v}$ is the symmetrized velocity gradient tensor.
We parameterize the magnitude of kinematic viscosity $\nu$ using a modified form 
of the $\alpha$-prescription 
\begin{equation}
\nu=\alpha \, c_{\rm s} \, Z \, {\cal F}_{\alpha}(r), 
\label{kinematic}
\end{equation}
where $c_{\rm s}^2=\gamma {\cal P}/\Sigma$ is the square of effective sound speed
calculated at each time step from the model's known ${\cal P}$ and $\Sigma$. The 
function ${\cal F}_{\alpha}(r)=
2 \pi^{-1} \tan^{-1}\left[(r_{\rm d}/r)^{10}\right]$ is a modification to the usual 
$\alpha$-prescription that guarantees that the turbulent viscosity operates 
only in the disk and quickly reduces to zero beyond the disk radius $r_{\rm d}$.
In this paper, we use a spatially and temporally 
uniform $\alpha$, with its value set to $5\times 10^{-3}$ to take into account
mass and angular momentum transport via such mechanisms as the magnetorotational 
instability. Transport of mass and angular momentum via gravitational instability 
is self-consistently taken into account via solution of the Poisson equation
for the gravitational potential of the disk and envelope.

The radiative cooling $\Lambda$ in equation~(\ref{energ}) is determined using the diffusion
approximation of the vertical radiation transport in a one-zone model of the vertical disk 
structure \citep{Johnson03}
\begin{equation}
\Lambda={\cal F}_{\rm c}\sigma\, T^4 \frac{\tau}{1+\tau^2},
\end{equation}
where $\sigma$ is the Stefan-Boltzmann constant, $T$ is the midplane temperature of gas, 
and ${\cal F}_{\rm c}=2+20\tan^{-1}(\tau)/(3\pi)$ is a function that 
secures a correct transition between the optically thick and optically thin regimes. 
We use frequency-integrated opacities of \citet{Bell94}.
The heating function is expressed as
\begin{equation}
\Gamma={\cal F}_{\rm c}\sigma\, T_{\rm irr}^4 \frac{\tau}{1+\tau^2},
\end{equation}
where $T_{\rm irr}$ is the irradiation temperature at the disk surface 
determined by the stellar and background black-body irradiation as
\begin{equation}
T_{\rm irr}^4=T_{\rm bg}^4+\frac{F_{\rm irr}(r)}{\sigma},
\label{fluxCS}
\end{equation}
where $T_{\rm bg}$ is the uniform background temperature (in our model set to the 
initial temperature of the natal cloud core $T_{\rm init}=10$~K)
and $F_{\rm irr}(r)$ is the radiation flux (energy per unit time per unit surface area) 
absorbed by the disk surface at radial distance 
$r$ from the central star. The latter quantity is calculated as 
\begin{equation}
F_{\rm irr}(r)= \frac{L_\ast}{4\pi r^2} \cos{\gamma_{\rm irr}},
\end{equation}
where $\gamma_{\rm irr}$ is the incidence angle of 
radiation arriving at the disk surface at radial distance $r$.

The stellar luminosity $L_\ast$ is the sum of the accretion luminosity $L_{\rm \ast,acr}=G M_\ast \dot{M}/2
R_\ast$, arising from the gravitational energy of accreted gas, and
the photospheric luminosity $L_{\rm \ast,ph}$ due to gravitational compression and deuterium burning
in the stellar interior. The stellar mass $M_\ast$ and accretion rate onto the star $\dot{M}$
are determined self-consistently during numerical simulations using the amount of gas passing through
the sink cell. The stellar radius $R_\ast$ is calculated using an approximation formula of \citet{Palla91},
modified to take into account the formation of the first molecular core.
The photospheric luminosity $L_{\rm \ast,ph}$ is taken from the pre-main 
sequence tracks for the low-mass stars and BDs calculated by \citet{DAntona97}. 
More details on the numerical code are given in \citet{VB2010a}.

\subsection{Initial conditions in pre-stellar cores}
We considered two limiting cases to describe the initial distribution of the gas surface density 
$\Sigma$ and angular velocity $\Omega$ in the pre-stellar cores. The first distribution, taken 
from \citet{Basu97}, is typical of pre-stellar cores formed as a result of the slow expulsion 
of magnetic field due to ambipolar diffusion, with the angular
momentum remaining constant during axially-symmetric core compression
\begin{equation}
\Sigma={r_0 \Sigma_0 \over \sqrt{r^2+r_0^2}}\:,
\label{dens}
\end{equation}
\begin{equation}
\Omega=2\Omega_0 \left( {r_0\over r}\right)^2 \left[\sqrt{1+\left({r\over r_0}\right)^2
} -1\right].
\label{omega}
\end{equation}
Here, $\Omega_0$ and $\Sigma_0$ are the angular velocity and gas surface
density at the disk center and $r_0 =\sqrt{A} c_{\rm s}^2/\pi G \Sigma_0 $
is the radius of the central plateau, where $c_{\rm s}$ is the initial sound speed in the core. 
The gas surface density distribution described by equation~(\ref{dens}) can
be obtained (to within a factor of unity) by integrating the 
three-dimensional gas density distribution characteristic of 
Bonnor-Ebert spheres with a positive density-perturbation amplitude A \citep{Dapp09}.
In all models the value of $A$ is set to 1.2, except for model~4 for which $A$=3.3.

The second set of initial conditions described by $\Omega$=const and $\Sigma$=const 
represent the other 
limiting case suggested in \citet{Boss01}. Cores that can be described, to a first degree of accuracy,
by spatially constant $\Sigma$ and $\Omega$
can form via gravitational fragmentation of filamentary structures, 
which are often encountered in numerical hydrodynamics simulations of the turbulent
fragmentation of giant molecular clouds. Another possible mechanism for the formation of
such cores may be the planar compression of pre-stellar condensations by shocks and UV radiation
of massive stars.

\subsection{Tracking the fragments}
\label{track}
Motivated by the absence of sink particles in our grid-based code,
we have designed a fragment-tracking algorithm which allows us to follow the trajectory 
of the fragments and calculate their physical parameters.
The most straightforward way to identify fragments in the disk 
is to set a threshold gas surface density $\Sigma_{\rm crit}$
which would help to distinguish between the fragments and the rest of the disk.
We however quickly found out that setting a single value of $\Sigma_{\rm crit}$, independent of radial
distance, did not work because too small a value for $\Sigma_{\rm crit}$ might result in spiral 
arms in the inner disk being identified as fragments. 
We therefore need to define a radially varying value of 
$\Sigma_{\rm crit}$ which is greater at smaller radii and vice versa.

Noting that the fragments are usually characterized by {\it peak} surface densities
that are greater than the local azimuthally averaged surface density $\overline\Sigma$ 
\citep{Vor2010,Vor2011}, the threshold gas surface density can be defined as 
$\Sigma_{\rm crit}=C \overline\Sigma$, where $C$ is an empirically determined constant.
 Numerical simulations show that the azimuthally averaged gas surface density in
gravitationally unstable disks declines with radius as 
${\overline\Sigma}=\overline{\Sigma}_0 (r_0/r)^{1.5}$ \citep[e.g.][]{Vor2010,Rice10}, 
where $\overline{\Sigma}_0$ is the typical azimuthally averaged density at $r_0$.  
For gravitationally unstable disks, $\overline\Sigma_0=20-50$~g~cm$^{-2}$ at 
$r\approx 100$~AU \citep{Vor2011}.
We therefore have chosen the following expression for the threshold density
\begin{equation}
\Sigma_{\rm crit} = \Sigma_{100} \left( {100~\mathrm{AU} \over r} \right)^{1.5}.
\label{critdens}
\end{equation}
The best value for threshold surface density at $r=100$~AU is found by trial
and error method to be $\Sigma_{100}=200$~g~cm$^{-2}$. Note that this value is 
greater than the mean gas surface density $\overline\Sigma_0$ by at least a factor of several.
The local maxima in the gas surface density that exceed $\Sigma_{\rm crit}$ 
usually represent the true fragments rather than the local maxima in the disk and/or spiral arms.
Although we may occasionally miss some of low-density fragments, this
does not invalidate our main conclusions.

After the radial $r_{\rm c}$ and angular $\phi_{\rm c}$ coordinates of the local maximum
representing the center of the fragment have 
been identified on the computational mesh, we determine the neighbouring cells that
belong to the fragment by imposing the following two conditions on the gas pressure $\cal P$
and gravitational potential $\Phi$
\begin{eqnarray}
\label{pres}
{\partial {\cal P} \over \partial r^\prime} &+& {1 \over r^\prime}{\partial {\cal P} \over \partial \phi^\prime} <0, \\
{\partial \Phi \over \partial r^\prime} &+& {1 \over r^\prime}{\partial \Phi \over \partial \phi^\prime} >0,
\label{grav}
\end{eqnarray}
where $r^\prime=r-r_{\rm c}$ and $\phi^\prime=\phi-\phi_{\rm c}$. The first condition 
mandates that the fragment must be pressure supported, with a negative pressure gradient
with respect to the center of the fragment. The second condition requires that the fragment
is kept together by gravity, with the potential well being deepest at the center of the fragment.
A substantial support against gravity may be provided by rotation but we assume that
this does not invalidate our criteria, i.e., no fragments assume a torus shape.

In practice, we start from the grid cell corresponding to the center of the fragment 
and proceed in eight directions (along the four coordinate directions and 
also at median angles to them)
until at least one of the above mentioned criteria is violated in every direction.
This procedure helps to identify an approximate shape of the fragment. 
We then check all the remaining grid cells that are encompassed by this octahedral shape 
and retain only those that meet both criteria~(\ref{pres}) and (\ref{grav}).
In addition, we filter out cells with the gas surface
density lower than that defined by equation~(\ref{critdens}), even if these cells 
still fulfil both criteria. We found that such cells are likely to belong 
to the circum-fragment disk rather than to the fragment itself. 
The cells that belong to the fragment 
are later utilized to calculate the mass and Hill radius and also the gravitational torque exerted on the fragment,
while the cell corresponding to the center of the fragment is used to calculate the trajectory of the
fragment. The characteristics of the fragments thus found  depend somewhat
on the adopted value of $\Sigma_{\rm 100}=200$~g~cm$^{-2}$. However, this dependence
is not critical and, for instance, the estimated masses of the fragments change by only 
about $10\%$ if $\Sigma_{\rm 100}$ is varied by a factor of two.

\section{Fragmentation and survival of fragments}
\label{fragmentation}
We have considered the evolution of 60 initial pre-stellar cores with
masses $M_{\rm c}$ ranging from 0.1~$M_\odot$ to 2.0~$M_\odot$ and ratios of rotational 
to gravitational energy $\beta$ lying between 0.27\% and 2.2\%. We have also varies the 
magnitude of the initial positive density perturbation $A$ in the 1.2--3.3 limits
and considered cores with distinct initial radial profiles of $\Sigma$ and $\Omega$.

For each model we ran our fragment-tracking algorithm at various evolution times
in order to identify models that experienced disk fragmentation. 
We found that most of the models showed disk fragmentation but at the same time 
failed to produce wide-orbit companions.
The majority of the fragments were torqued into the inner disk 
region and through the sink cell 
(6~AU) producing mass accretion and luminosity bursts similar in magnitude
to those of FU-Ori-type objects \citep{VB06,VB2010a},
while the remaining few were ejected from the disk into the intracluster medium 
via many-body gravitational interaction \citep{BV12} or dispersed via tidal torques 
\citep{Vor2011a,Boley10}. \citet{Boley10} and \citet{Zhu2012} also found support for clump-driven
FUor events.
Only 6 models out of 60 have shown the formation of stable companions in wide orbits, 
with their mass ranging from 3.5 to 43 Jovian masses. 
Our obtained survival probability is even lower than that of \citet{Zhu2012} (3 out of 13), though
they followed the evolution of the fragments for a significantly shorter time period. 

\begin{figure}
  \resizebox{\hsize}{!}{\includegraphics{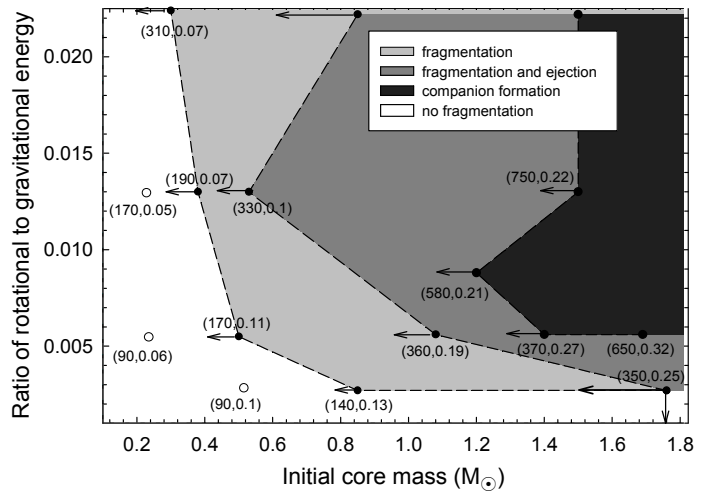}}
  \caption{Phase space of $\beta$, the initial ratio of rotational to gravitational energy,
  vs. core mass $M_{\rm c}$.  The solid/open circles correspond to models with/without 
  disk fragmentation.  The black region
  is the region where the formation of sub-stellar- or planetary-mass companions on wide orbits 
  is found in this study. The dark-shaded region is the region in which both fragmentation and
  ejection events may occur, and the light-shaded region is the one where only
  fragmentation events occur. The white area marks the region where no fragmentation is
  observed. Arrows illustrate uncertainties associated with a
  coarse grid of models and indicate models in the $\beta:M_{\rm c}$ phase space 
  that might have shown (with a 50\% probability) 
  disk fragmentation (ejection/companion formation) had we considered
  a finer grid of models. Each pair of data in parenthesis indicate the mean disk radius and
  mass for the corresponding model. }
  \label{fig1}
\end{figure}

Figure~\ref{fig1} illustrates our findings on the phase space of $\beta$ versus $M_{\rm c}$ 
covered in our modeling. The light-shaded area defines the region where disk fragmentation takes place,
while the dark-shaded area outlines the region where both fragmentation and ejection of the fragments
may occur. The data related to disk fragmentation and ejection of the fragments were taken 
from numerical hydrodynamics simulations of \citet{BV12}. In the present study
we added several models that revealed the survival of the fragments. The total number of models 
used in the current study amounts to 60 but only 16 key models 
(those lying at the boundary of the considered regions) 
have been shown in Figure~\ref{fig1} to avoid overcrowding.

Previous numerical and analytical studies of protostellar disk
evolution demonstrated that gravitational fragmentation is possible at distances 
$\ga 50$~AU \citep[e.g.][]{Clarke09,Stamatellos09,Boley09,Rice10} and in disks 
with mass $\ga 0.1~M_\odot$ \citep[e.g.][]{Rice03,Mayer07}. 
Disk masses and radii in our models can be calculated
by disentangling the disk and infalling core
on the computational mesh. This is however not an easy task.
We do this in practice by 
first constructing the azimuthally averaged gas surface density profiles
$\overline{\Sigma}$ and then applying
a threshold value of $\overline{\Sigma}_{\rm d2e}=0.5$~g~cm$^{-2}$ between the disk and
the core. We also use the radial gas velocity profile to see where 
the infalling envelope lands onto the disk \citep[see][for details]{Vor2011}. 
The disk radius estimates are further complicated by substantial outward 
excursions (or even complete ejections) of some fragments. 
It is not always clear if the fragment belongs to the 
circumstellar disk or it has already detached from it and 
should be treated as an external companion (see e.g. Figure~\ref{fig2}). 
We assume that the fragment belongs to the disk if $\overline{\Sigma}$ between
the fragment and the disk does not drop below 0.05~g~cm$^{-2}$, a value
typical for the interface between the disk and the external environment \citep{Vor2011}.
In the opposite case, the fragment is supposed to have detached from the natal
disk and is treated as a separate entity.

Our choice of a threshold of 0.5~g~cm$^{-2}$ is motivated
by the fact that the resulting distribution of disk radii peaks
between $10^2$ and $10^3$~AU; adopting an order of magnitude lower
threshold would shift the entire distribution up by about a factor
of 2--3 towards bigger values. 
Sizes of embedded disks are very poorly constrained by
observations and are typically assumed to be on the order of
100 AU or less based on simple centrifugal radius arguments. 
In reality, however, gravitational and viscous transport of mass and angular momentum 
will cause disks to spread to sizes greater than the corresponding centrifugal radii. 
We note that there are some limited observations supporting the existence of
large protostellar disks \citep[e.g.][]{Enoch09,Jorgensen09}. 
Ultimately, the correct
threshold to adopt in order to disentangle the disk and core in
the simulations will remain uncertain until the masses and sizes
of protostellar disks are better constrained from observations.

Each pair of data in Figure~\ref{fig1} indicates the mean disk radius and mass for 
the corresponding model marked with circles. 
The mean values are calculated
by time-averaging the instantaneous values over the duration of the Class~I
phase, in which disk fragmentation is usually most vigorous.
It is evident that the disk mass and radius {\it both} have to exceed 
some threshold values in order for the disk to fragment. 
In particular, models with higher $\beta$ experience disk fragmentation 
at lower disk masses then their low-$\beta$ counterparts.
The minimum mean disk mass at which fragmentation can take place according to our modeling
is $\overline{M}_{\rm d}^{\rm fr}=0.07~M_\odot$ for $\beta\ga 1.3\%$. The latter value may
increase by about a factor of 2 for models with lower $\beta$.  
We take into account here the uncertainty illustrated by the errors in Figure~\ref{fig1}
associated with a coarse grid of models.
Our found values of $\overline{M}_{\rm d}^{\rm fr}$ are in reasonable agreement with 
previous estimates, 0.1~$M_\odot$ \citep[e.g.][]{Rice03,Mayer07}.
The minimum radius of a disk experiencing fragmentation in our models is 
$\overline{R}_{\rm d}^{\rm fr}=140$~AU for $\beta\approx 0.3\%$. 
The latter value may increase by up to a factor of 2
for models with higher $\beta$.  Our derived values of $\overline{R}_{\rm d}^{\rm fr}$ 
are difficult to compare with other studies of disk fragmentation 
because the latter tend to provide the minimum 
radial distance at which fragmentation can take place ($r_{\rm fr}$) 
and not the minimum disk radius ($\overline{R}_{\rm d}^{\rm fr}$) that is required
for disk fragmentation to occur.
Obviously, $\overline{R}_{\rm d}^{\rm fr}$ must be grater than $r_{\rm fr}$ and our simulations suggest
that $\overline{R}_{\rm d}^{\rm fr}=(3-6) r_{\rm fr}$.

The black area in Figure~\ref{fig1} marks the region where the formation
of GP and/or BD companions on wide orbits is found in the present work. 
None of our models have shown the formation of wide-orbit companions for $\beta\ga 1.5\%$
but we attribute this to a rather narrow sampling of models at this region.
Therefore, we extrapolated the companion-forming domain to $\beta \ga 1.5\%$ assuming 
the same likelihood for the formation of wide-orbit companions as for models with lower $\beta$.
The companion-formation domain is notably
narrower than both the fragmentation and fragmentation plus ejection domains,
particularly as far as the initial core mass is concerned.
It appears that $M_{\rm c}$ has to be greater than 1.2~$M_\odot$ and the
ratio of rotational to gravitational energy has to exceed 0.5\% in order to
make the formation of wide-orbit companions possible.
Moreover, the minimum mean disk mass at which the formation of wide-orbit companions is 
found in our modeling is $\overline{M}_{\rm d}^{\rm c.f.}=0.21~M_\odot$ for models
with $\beta\ga0.8\%$. The value of $\overline{M}_{\rm d}^{\rm c.f.}$ 
may increase somewhat for lower-$\beta$ models. 
The minimum mean disk radius that is required for the formation of wide-orbit companions
is $\overline{R}_{\rm d}^{\rm c.f.}=370$~AU and this value increases with 
increasing $\beta$. 
Our found values of the minimum disk radius and mass both depend on the adopted threshold.
For instance, adopting a factor of five lower density threshold $\overline{\Sigma}_{\rm d2e}=0.1$~g~cm$^{-2}$
would yield a factor of 1.2 greater minimum disk mass $M_{\rm d}^{\rm c.f.}=0.25~M_\odot$
and a factor of 1.3 greater minimum disk radius $\overline{R}_{\rm d}^{\rm c.f.}=470$~AU.

It is not clear why models with $M_{\rm c}<1.2~M_\odot$ fail to produce wide-orbit companions.
Large masses and radii of protostellar disks in companion-forming models must certainly be factors that
help wide-orbit companions to survive. 
Another likely reason why low-$M_{\rm c}$ models fail to form wide-orbit 
companions is that disk fragmentation in these models is mostly confined to the
embedded phase of star formation and is sustained 
by continuing mass loading from an infalling envelope.
A fragment may escape inward migration if the net torque exerted on it is not negative. 

To a first order of accuracy, the net torque acting on the fragment in a
gravitationally unstable disk can be expressed as the sum of the gravitational 
torques from the inner ${\cal T}_{\rm in}$ and outer ${\cal T}_{\rm out}$ 
parts of the disk with respect to the current position of the fragment. 
Spiral arms and other fragments are the main contributors to the total torque. Due to the trailing nature
of the spiral arms in gravitationally unstable
protostellar disks, ${\cal T}_{\rm in}$ is usually positive and ${\cal T}_{\rm out}$ 
is usually negative.  The condition for the fragment to avoid inward migration 
can then be written as
\begin{equation}
{d L \over d t } = {\cal T}_{\rm in} + {\cal T}_{\rm out} \ge 0,
\label{nettorque}
\end{equation}
where $L$ is the angular momentum of the fragment. Evidently, the fragment
will stay in the disk for as long as $|{\cal T}_{\rm out}|\le {\cal T}_{\rm in}$.

Even if the fragment forms near the disk outer edge where inequality~(\ref{nettorque}) 
is likely to be fulfilled initially,
it can start migrating inward during the subsequent evolution due to continuing mass loading from
the parental core. The mass infall onto the disk is a double-edged sword effect: 
it promotes disk fragmentation by increasing the disk mass 
\citep[e.g.][]{VB06,VB2010a,Kratter10a} but it also deposits a sub-Keplerian material 
at/near the disk outer region \citep[e.g.][]{VD10}.
The latter acts to increase $|{\cal T}_{\rm out}|$ as the mass near the disk outer edge 
accumulates so that the fragment starts to migrate inward when $|{\cal T}_{\rm out}|$
becomes greater than ${\cal T}_{\rm in}$. In addition,
a sub-Keplerian material falling onto the disk can exert a torque onto 
the fragment and push the later in towards the star.

On the contrary, models with $M_{\rm c}\ga 1.2~M_\odot$ are usually characterized by disks
that are sufficiently massive and large to experience fragmentation not only 
in the embedded phase and 
but also in the T Tauri phase when mass loading onto the disk diminishes (see Figures~\ref{fig2}
and \ref{fig8}). 
When formed at the disk outer regions where inequality~(\ref{nettorque}) is fulfilled, 
fragments in these models have chances to open a gap in the disk and settle on quasi-stable, wide orbits.
The main conclusion drawn from 
our parameter space study is that disk fragmentation is {\it not} sufficient to
guarantee the formation of GP or BD companions on wide orbits.

In this paper, we used a spatially and temporally uniform $\alpha=5\times10^{-3}$.
This choice is based on the work of \citet{VB09,VB09b}, who found that models
with $\alpha=10^{-2}$ reproduce well the slope of the 
mass accretion rate--stellar mass relation for young 
brown dwarfs and low-mass stars ($\sim 1.0$~Myr~old), though slightly overpredicting the mean
accretion rates.
Greater value of $\alpha$ act to weaken the strength of gravitational instability
due to an overall increase
in the efficiency of mass transport and the corresponding decrease in the total disk mass. 
However, large values of $\alpha$ ($\ga 10^{-1}$) destroy circumstellar disks during less 
than 1.0~Myr of evolution and are thus inconsistent with mean disk lifetimes
on the order of 2–-3 Myr. 
We therefore conclude that a moderate increase in 
the adopted value of $\alpha$ (by a factor of several) can shift
the fragmentation boundary in Figure~\ref{fig1} towards higher values of 
$M_{\rm c}$ and $\beta$ but is not expected to shut off disk fragmentation completely.

Below, we present six models that have shown the formation of 
GP/BD embryos on wide, quasi-stable orbits via disk gravitational fragmentation.
We pay most attention to models~1--3 and
provide only main results for the other three models. 
The parameters of the models are listed in Table~\ref{table1}.

\begin{table}
\begin{center}
\caption{Model parameters}
\label{table1}
\renewcommand{\arraystretch}{1.5}
\begin{tabular*}{\columnwidth}{ @{\extracolsep{\fill}} c c c c c c}
\hline \hline
model & $M_{\rm c}$ & $\beta$ & $\Sigma_0$ & $\Omega_0$ & $r_0$ \\
\hspace{1cm} &  ($M_\odot$)   & (\%) & (g~cm$^{-2}$) & (s$^{-1}$) & (AU)  \\ [0.5ex]
\hline \\ [-2.0ex]
1 & 1.7 & 0.56 & $3.3\times10^{-2}$ & $2.4\times10^{-14}$ & 3770 \\
2$^{\dagger}$ & 1.2 & 0.88 & $1.3\times 10^{-2}$ & $1.3\times10^{-14}$ & -- \\
3 & 1.5 & 0.56 & $3.1\times 10^{-2}$ & $2.6\times10^{-14}$ & 3430 \\
4 & 1.4 & 0.56 & $1.1\times 10^{-1}$ & $6.0\times10^{-14}$ & 1890 \\
5 & 1.55 & 1.27 & $3.6\times 10^{-2}$ & $4.0\times10^{-14}$ & 3430 \\
6 & 1.4 & 0.56 & $4.0\times 10^{-2}$ & $3.9\times10^{-14}$ & 3090 \\
\hline
\end{tabular*}
\end{center}
\medskip
{$^\dagger$ Models with spatially constant gas surface density $\Sigma\equiv\Sigma_0$ and
angular velocity $\Omega\equiv\Omega_0$.}
\end{table}

\section{Formation of a planetary-mass companion}
\label{planet}
\begin{figure}
  \resizebox{\hsize}{!}{\includegraphics{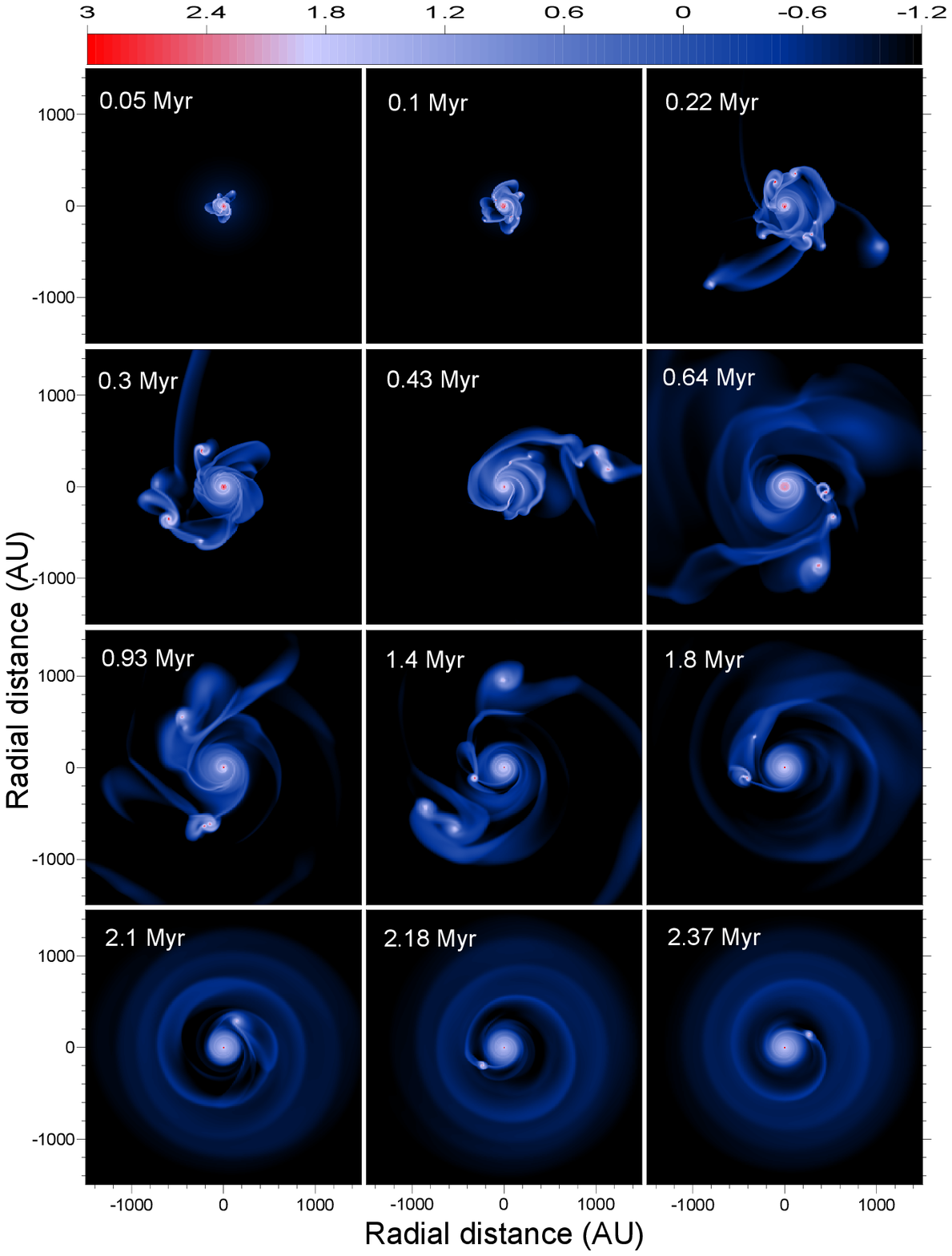}}
  \caption{Gas surface density distribution in model~1 shown at various times 
  since the formation of the central protostar. Only the inner $3000\times3000$~AU box is shown,
  the total computational region extends to 22000~AU. The scale bar is in log~g~cm$^{-2}$.
  Note a fragment on a stable orbit in the bottom row.}
  \label{fig2}
\end{figure}

In this section, we describe the formation of an 11-Jupiter-mass companion around
a 1.2~$M_\odot$ star. Figure~\ref{fig2} presents a series of the gas surface density images 
showing the evolution of the disk in model~1 starting soon after the formation of the central object
($t=0.05$~Myr) and ending in the T Tauri phase ($t=2.37$~Myr). The box size is $3000$~AU 
on each side and represents a small subregion of the overall computational domain.  
The time elapsed after the formation the central protostar is shown in each image 
and the minimum gas surface density plotted in the Figure is 0.06~g~cm$^{-2}$ or -1.2 
in the log units.  

The forming disk is gravitationally unstable and 
first fragments start to appear in the disk as early 
as at 50~kyr after the formation of the central star. Gravitational 
perturbations from spiral arms and massive fragments cause significant radial motions in the disk.
As a result, the disk appears as a somewhat chaotic structure with dense filaments connecting
the fragments. Nevertheless, a near Keplerian
rotation can be retrieved after azimuthal averaging.
Fragmentation is predominantly concentrated to the intermediate and outer disk regions,
a consequence of the mass infall and stellar irradiation,
and no fragmentation is evident at $r \la 50-100$~AU.

One may notice from Figure~\ref{fig2} that the number of fragments is varying with time, 
indicating that the fragments may be tidally destroyed or otherwise lost by the disk. 
For instance, fragments may
migrate inward onto the star \citep{VB2010a,Machida2011} or get ejected from the disk 
into the intracluster medium \citep{BV12}. As a result of these migration/ejection/destruction  
processes, only one fragment survives after 1.7~Myr of the disk evolution. 
The bottom row in Figure~\ref{fig2} reveals 
the typical picture with the surviving fragment opening a gap and inducing spiral waves
in the disk. 
The fragment itself is connected with the inner and outer disks by a wake of enhanced
surface density which trails/leads the fragment outside/inside its orbit.

\begin{figure}
  \resizebox{\hsize}{!}{\includegraphics{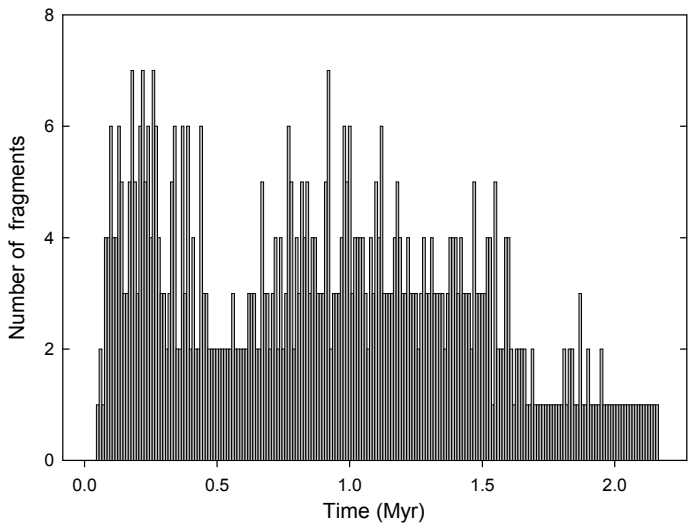}}
  \caption{Number of fragments vs. time in model~1. The number of fragments at a given time instant
  is calculated using the fragment tracking algorithm described in \S~\ref{track}.
  An increase in the number of fragments shows recent fragmentation,
  and a decrease shows recent destruction/accretion of the fragments.}
  \label{fig3}
\end{figure}

Figure~\ref{fig3} presents the number of fragments in the disk at a given time instant.
The number of fragments varies with time and a maximum value
($N_{\rm fr}^{\rm max}=7$) is reached at $t\approx0.2$~Myr and $t\approx0.9$~Myr. Between
these two maxima, a local minimum with just two fragments in the disk occurs 
at $t\approx0.5$~Myr. The disk mass at the end of the embedded phase (t=0.65~Myr) 
is $\approx 0.3~M_\odot$, sufficient to sustain fragmentation in the disk for another
0.8~Myr.

\begin{figure}
  \resizebox{\hsize}{!}{\includegraphics{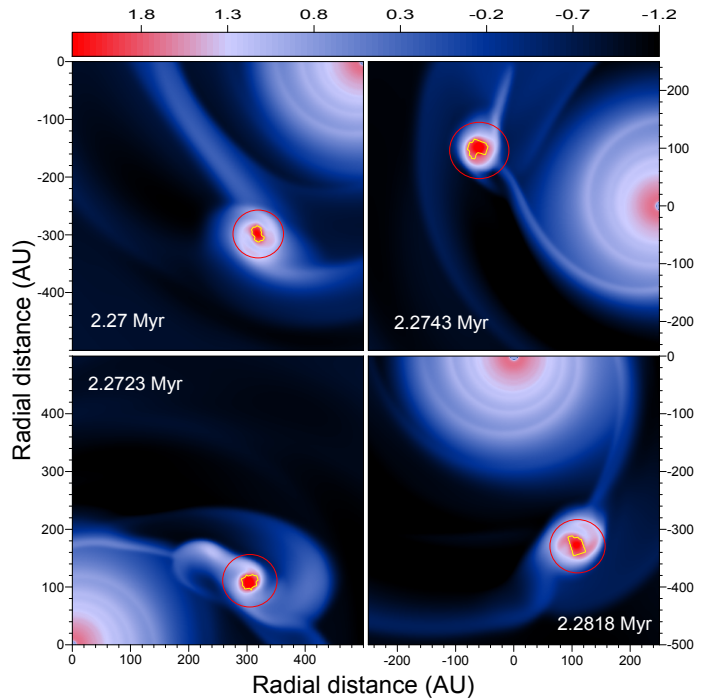}}
  \caption{Zoomed-in view on the surviving fragment in model~1 at four time instances. 
  The color image shows the gas surface density in log~g~cm$^{-2}$, the yellow curve 
  outlines the fragment and the red circles marks the Hill radius.}
  \label{fig4}
\end{figure}

Figure~\ref{fig4} shows four zoomed-in images of the surviving fragment taken
during a time period of 1180~yr covering just two orbital periods of the fragment. 
The box size is $500\times500$~AU.
The fragment is outlined by the yellow curve, which is found using the tracking algorithm
described in Section~\ref{track}. The red circle represents the corresponding Hill radius
of the fragment calculated as 
\begin{equation}
R_{\rm H}= r_{\rm f} \left( {1\over 3} {M_{\rm f} \over M_\ast + M_{\rm f}} \right)^{1/3},
\end{equation}
where $M_\ast$ is the stellar mass, $r_{\rm f}$ is the orbital distance
of the fragment and $M_{\rm f}$ is the mass of the fragment confined
within the yellow curve. 
A mini-disk with a developed two-armed spiral structure can be seen around the fragment
in the lower-right image. 

\begin{figure}
  \resizebox{\hsize}{!}{\includegraphics{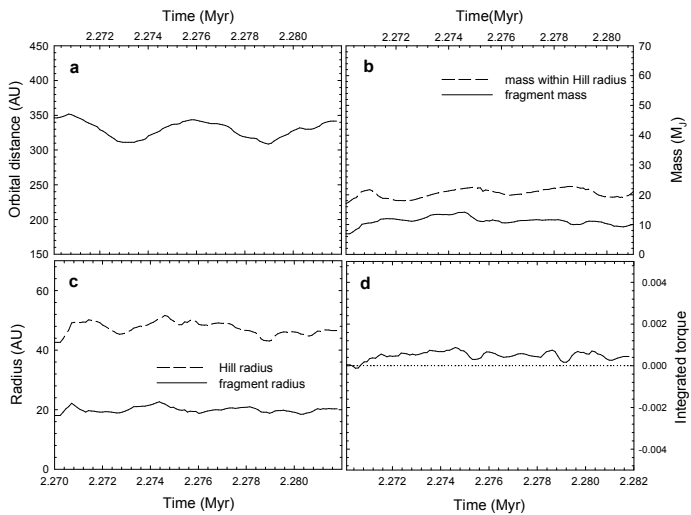}}
  \caption{Characteristics of the surviving fragment in model~1: (a) orbital distance,
  (b) mass of the fragment (solid line) and that confined within the Hill radius
  (dashed line), (c) radius of the fragment (solid line) and the Hill radius (dashed line),
  and (d) integrated gravitational torque acting on the fragment 
  in units of $8.6\times 10^{40}$~g~cm$^2$ s$^{-2}$.}
  \label{fig5}
\end{figure}

We used our fragment tracking algorithm to calculate physical properties of the
surviving fragment. 
Figure~\ref{fig5} presents (a) the orbital distance of the fragment $r_{\rm f}$,
(b) the mass of the fragment $M_{\rm f}$ and the mass confined within 
the Hill radius $M_{\rm H}$ (solid and dashed lines, respectively), (c)
the radius of the fragment $R_{\rm f}$ and the Hill radius $R_{\rm H}$ (solid and dashed lines, respectively),
and (d) the integrated gravitational torque acting on the fragment $\cal T$ (solid line).
The latter quantity is calculated as the sum of all individual torques 
$\tau=-m(r,\phi)\partial \Phi/\partial \phi$ acting on the fragment,  where $m(r,\phi)$ is the 
gas mass in a cell with polar coordinates $(r,\phi)$ and $\Phi$ is the corresponding 
gravitational potential.   The dotted line marks the zero torque. 

The orbital distance of the fragment varies in the 308--352~AU limits. The mean
distance from the central star is $\overline{r}_{\rm f}=330$AU and eccentricity
of the orbit is $\epsilon\approx 0.07$.
Such low eccentricity orbits are typical for companions formed by disk gravitational instability
\citep{VB2010b, Boss12}.
The mass of the fragment varies in the 7--14~$M_{\rm J}$ limits. This wide scatter 
reflects either imperfections in the fragment tracking mechanism or continuous perturbations 
imposed onto the fragment by the circumfragment disk and spiral density wake (or both effects). 
These perturbations, however, do not lead to the
fragment dispersal at least on time scales of our numerical simulations.
The mean mass of the fragment calculated over two orbital periods shown in Figure~\ref{fig5} 
is $\overline{M}_{\rm f}$=11~$M_{\rm J}$, which is still
in the planetary mass regime. However, the mean mass confined within the Hill radius is 
$\overline{M}_{\rm H}=20.5~M_{\rm J}$, which implies that the fragment may accumulate some
more material in the course of the evolution as it cools and contracts into
planetary-sized object. On the other hand, if the fragment could not get rid of 
angular momentum, most of the material in the Hill
sphere would ultimately land onto a circumfragment disk and the fragment
may remain the planetary-mass regime.

Figure~\ref{fig5}c shows the radius of the fragment and the Hill radius. 
The mean Hill radius $\overline{R}_{\rm H}=47$~AU is greater than mean radius of the fragment
$R_{\rm f}=20$~AU by more than a factor of 2.
The mean scale height at the position of the planet is about $Z=40$~AU, somewhat
smaller than the Hill radius. According to \citet{Crida06} and \citet{Kley12}, 
the gap opening criterion can be written as
\begin{equation}
{3\over 4} {Z \over R_{\rm H}} + {50 \nu \over q r_{\rm f}^2 \Omega_{\rm f}} \la 1.0,
\label{gap}
\end{equation}
where $q=M_{\rm H} /M_\ast$ and $\Omega_{\rm f}$ is the orbital frequency of the fragment.
Substituting the corresponding mean values for $R_{\rm H}$ and $r_{\rm f}$ into 
equation~(\ref{gap}), noticing that $q=0.017$ for $M_\ast=1.2~M_\odot$ and 
$M_{\rm H}=20.5~M_{\rm J}$, and finally
calculating $\nu$ using equation~(\ref{kinematic}) and mean disk temperature of 15~K at 
$r_{\rm f}=330$~AU, we estimated the left-hand side of equation~(\ref{gap}) to 
be $\approx 0.9$, thus marginally satisfying the gap opening criterion. 
The future orbital dynamics of the fragment can be predicted using the
integrated gravitational torque acting on the fragment from the rest of the disk, $\cal T$, shown
in Figure~\ref{fig5}d.
Evidently, the torque is  mostly positive, implying outward migration, and its mean value is 
$\overline{{\cal T}}=4.9\times 10^{-4}$ in units of  $8.6\times 10^{40}$~g~cm$^2$ s$^{-2}$. 

We estimate the characteristic migration timescale using the following simple analysis.
A (small) change of the orbital distance $dr_{\rm f}$ of a fragment 
with mass $M_{\rm f}$ on a Keplerian orbit caused by a (small) change in the angular momentum 
of the fragment $dL$  can be written as $dr_{\rm f}= 2\, dL/M_{\rm f} v_{\rm f}$, 
where $v_{\rm f}=(GM_\ast/r_{\rm f})^{1/2}$ is the angular velocity of the fragment. 
The migration velocity of the fragment is then
\begin{equation}
v_{\rm mg}={dr_{\rm f}\over dt} = {2 {dL \over dt} \over M_{\rm f} v_{\rm f}}.
\end{equation}
Noticing that $dL/dt=\overline{\cal T}$, 
the characteristic migration time can be calculated as
\begin{equation}
t_{\rm mg}= {r_{\rm f} \over v_{\rm mg}}= {L \over 2 \overline{{\cal T}}}.
\label{migrate}
\end{equation}
Substituting the corresponding mean values for $r_{\rm f}$, $M_{\rm f}$ 
and $\cal T$ into equation~(\ref{migrate}), and also noticing that
$M_\ast=1.2~M_\odot$ in model~1, we estimated the migration timescale
to be of the order of 10~Myr. 

Since the fragment in model~1 opens a gap in the disk, 
a more appropriate estimate of the migration timescale may be that
given by the viscous diffusion time in the disk \citep{Lin86}
\begin{equation}
t_{\rm visc}={r_{\rm f}^2 \over \nu},
\end{equation}
which yields essentially the same timescale of 10~Myr.
The above estimates
show that the fragment will remain on a wide orbit for
a time period longer than the typical disk lifetime of 2--3~Myr \citep{Strom,Haisch},
implying that the fragment will finally turn into a massive GP or low-mass BD
on a stable orbit of the order of 400~AU from the central object.

\section{Formation of an intermediate-mass brown dwarf}
\label{browndwarf}
\begin{figure}
  \resizebox{\hsize}{!}{\includegraphics{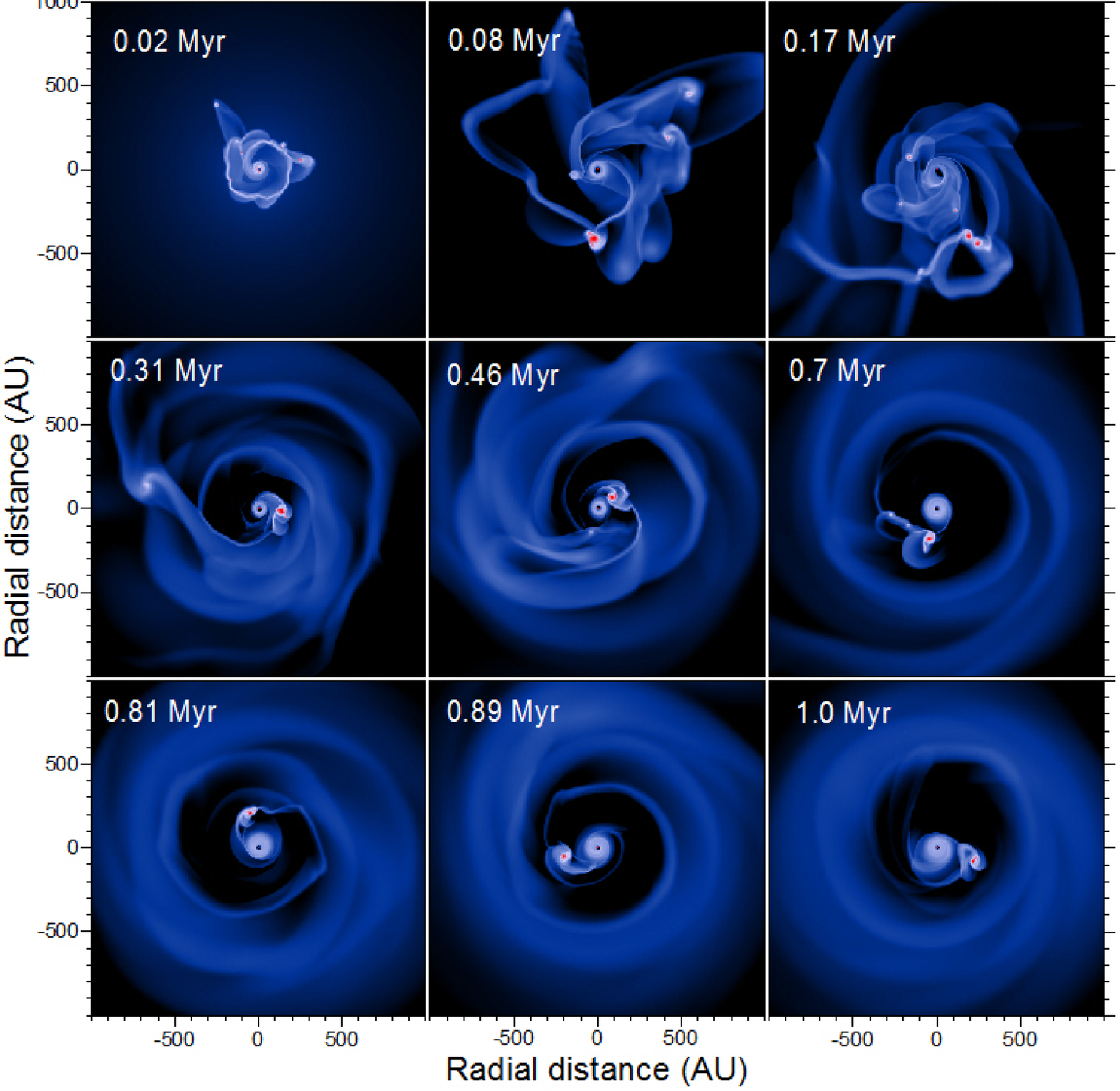}}
  \caption{Gas surface density distribution in model~2 shown at various times 
  since the formation of the central protostar. Only the inner $2000\times2000$~AU box is shown,
  the total computational region extends to 16000~AU. The scale bar is in log~g~cm$^{-2}$.
  Note a fragment on a stable orbit in the bottom row.}
  \label{fig6}
\end{figure}

In this section we describe the formation of a 43-Jupiter-mass BD
around a $0.9~M_\odot$ star. Figure~\ref{fig6} shows a series of images of the 
gas surface density (logarithmic in g~cm$^{-2}$) in model~2 for the inner 1000~AU
of our computational box.
The time elapsed since the formation of the central protostar is indicated in each image.
This model is characterized by initial gas surface density and angular velocity
in the pre-stellar core that are independent of radial
distance, i.e.,  $\Sigma=\mathrm{const}$  and $\Omega=\mathrm{const}$. 
As a consequence, model~2 has an elevated mass infall
onto the disk, stronger gravitational
instability and more vigourous fragmentation in the Class 0 phase \citep{Vor2012} 
than model~1, notwithstanding the fact that the prestellar core in the latter model is more massive.
Gravitational instability in model~2, fueled by intense mass loading from the envelope, 
is so strong that the disk has broken into massive clumps linked with each other 
by long and dense filaments. During the course of the evolution, most fragments have 
migrated onto the star
due to strong torques but one massive fragment manages to survive through the initial violent
stage and settles onto a quasi-stable orbit at around 0.3~Myr.

\begin{figure}
  \resizebox{\hsize}{!}{\includegraphics{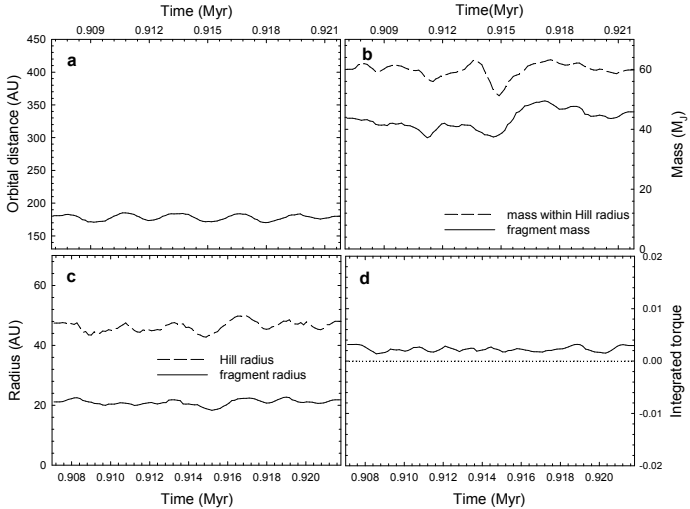}}
  \caption{Same as Figure~\ref{fig5} only for model~2.}
  \label{fig7}
\end{figure}

Figure~\ref{fig7} presents main characteristics of the surviving fragment during
approximately four orbital revolutions. The layout of the Figure is the same as that of 
Figure~\ref{fig5}. 
The orbital distance of the fragment varies in the 170--185~AU limits and the mean distance 
is $\overline{r}_{\rm f}=178$~AU.
The fragment is characterized by a low eccentricity orbit, $\epsilon\approx0.04$, somewhat smaller
than that of the fragment in model~1. The mean mass of the fragment is 
$\overline{M}_{\rm f}=43~M_{\rm J}$ and the 
mean mass contained within the Hill radius is $\overline{M}_{\rm H}=60~M_{\rm J}$. This implies that
the fragment will ultimately form an intermediate mass BD, perhaps surrounded by 
its own circum-BD disk. The mean radius of the fragment 
is $\overline{R}_{\rm f}=21$~AU and the mean
Hill radius is $\overline{R}_{\rm H}=46$~AU. The latter value is greater than the 
local scale height, $Z\approx24$~AU, and the gap opening criterion~(\ref{gap}) is satisfied.
In general, the fragment in model~2 appears to be in a more perturbed state than that
of model~1, perhaps, because the former is more massive, younger and less evolved.

The integrated gravitational torque acting on the fragment in model~2 is always positive and
is somewhat stronger than that of model~1, possibly due to a higher mass of the former. 
The migration timescale calculated using equation~(\ref{migrate})
is found to be of the order of 4~Myr, which is comparable to or even longer than the typical lifetime
of the disk. We conclude that the fragment in model~2 has good chances to survive migration and 
settle on a wide
orbit evolving ultimately into an intermediate-mass BD.

\section{Attempted formation of multiple companions}
\label{multiple}
Notwithstanding the fact that protostellar disks in models~1 and 2 exhibit multiple episodes
of gravitational fragmentation, 
only one fragment in each model has survived after 1.0~Myr of disk evolution.
This raises the question of whether
gravitational fragmentation can account for the formation 
of multi-companion systems similar to HR 8799 \citep{Marois08, Marois10},
which has four planetary-mass objects on orbits at 15--70~AU from the central star.
Below, we discuss this possibility.

\begin{figure}
  \resizebox{\hsize}{!}{\includegraphics{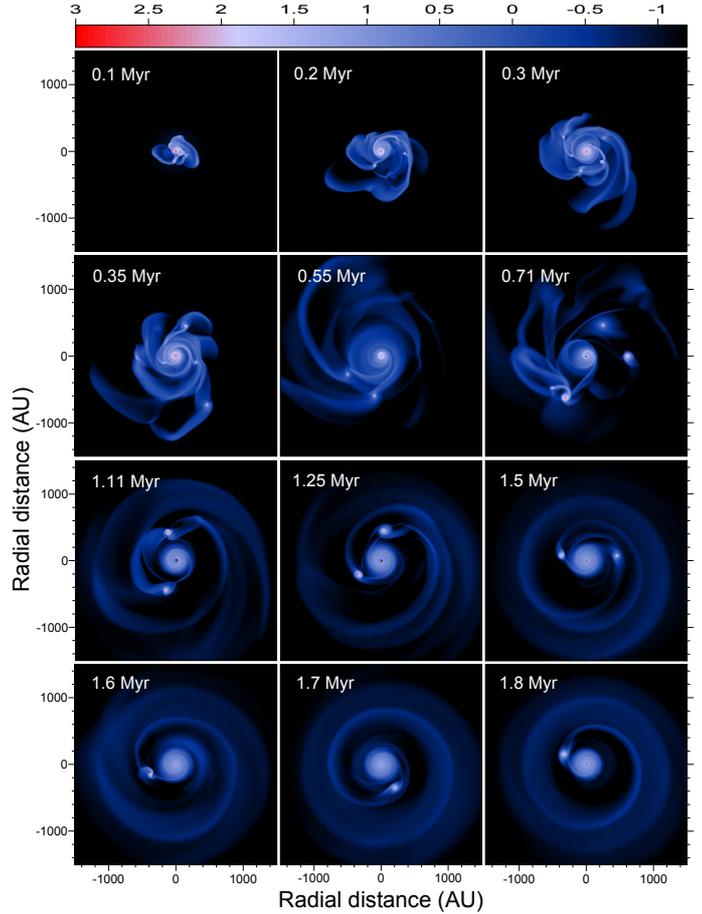}}
  \caption{Gas surface density distribution in model~3 shown at various times 
  since the formation of the central protostar. Only the inner $2000\times2000$~AU box is shown,
  the total computational region extends to 20000~AU. The scale bar is in log~g~cm$^{-2}$.
  Note two fragments on quasi-stable orbits in the third row of images. One of the fragments
  disperses at $t\approx1.55$~Myr and the other survives to the end of numerical simulations 
  ($t=1.8$~Myr).}
  \label{fig8}
\end{figure}

Figure~\ref{fig8} presents a series of images of the gas surface density in model~3.
The parameters of the model are listed in Table~\ref{table1} and the time after the formation
of the protostar is indicated in each image. The initial evolution of the disk is characterized by vigourous
fragmentation and several fragments at a time are usually present in the disk. By $t=1.1$~Myr only
two fragments survive and settle on quasi-stable orbits with only slightly different 
radial distances from the star but with a $160^\circ$ offset in azimuthal angle 
with respect to each other.
However, after orbiting in unison for about 0.45~Myr, one of the fragments
disperses at $t\approx1.55$~Myr. The dispersed fragments leaves a
a crescent-shaped density enhancement in the disk which can still be seen in Figure~\ref{fig8} at 
$t=1.6$~Myr. The other fragment survives to the end of
our numerical simulations ($t=1.8$~Myr).

\begin{figure}
  \resizebox{\hsize}{!}{\includegraphics{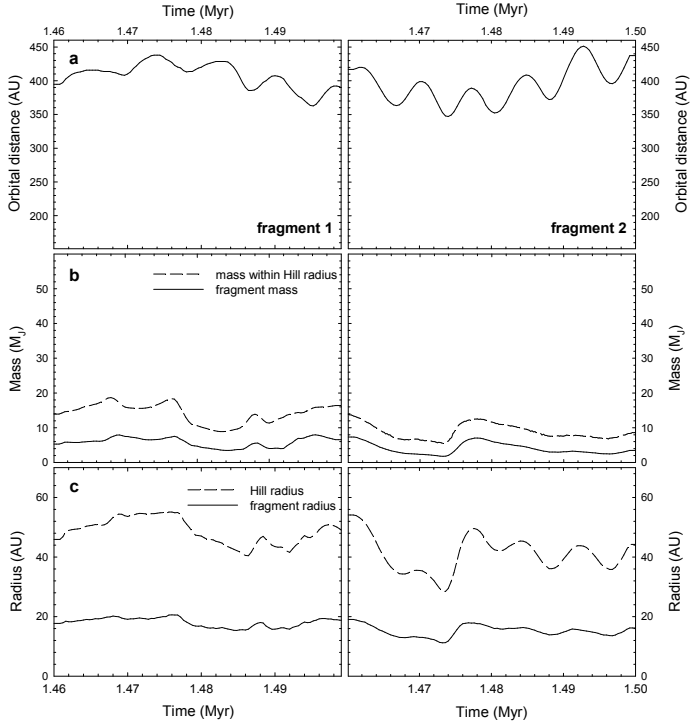}}
  \caption{Main characteristics of fragment~1 (left column) and fragment~2 (right column) 
  in model~3 before the least massive of them (fragment~2) has dispersed.
  In particular, Panel~a presents the orbital distance of the fragments,
  Panel~b---masses of the fragments (solid lines) and those confined within the Hill radii
  (dashed lines), and Panel~c---radii of the fragments (solid lines) and their Hill radii (dashed lines).}
  \label{fig9}
\end{figure}

Figure~\ref{fig9} shows main characteristics of the two surviving 
fragments during a time period of 1.46--1.5~Myr, i.e., before one of the fragments has 
dispersed. In particular, the left-hand and right-hand panels belong
to fragment~1 (most massive) and fragment~2 (least massive), respectively. 
Panels~(a) in Figure~\ref{fig9} present the orbital distance
of the fragments,  panels~(b)---masses of the fragments (solid lines) and masses 
contained within the Hill radii of each fragment (dashed lines), and
panels~(c)---radii of the  fragments (solid lines) and their Hill radii (dashed lines).

Both fragments move on orbits that are less stable than those in models~1 and 2,
perhaps due to continuing gravitational perturbation exerted on the fragments by spiral
density wakes excited by both fragments in the disk. The mean orbital distances
of fragment~1 and 2 are $\overline{r}_{\rm f,1}=407$~AU and $\overline{r}_{\rm f,2}=393$~AU, respectively.
The eccentricity of the orbits is also varying somewhat and the maximum eccentricity 
for fragments~1 and 2 are estimated to be approximately $\epsilon_1=0.05$ and $\epsilon_2=0.1$, respectively.
As a result, notable radial excursions are evident in the top panels of Figure~\ref{fig9}.
The masses of both fragments stay in the planetary-mass regime, though with
significant variations reflecting their highly perturbed state,  
and the mean masses of fragments~1 and 2 are $\overline{M}_{\rm f,1}=5.9$~$M_{\rm J}$ 
and $\overline{M}_{\rm f,2}$=4.0~$M_{\rm J}$, respectively.
The mean masses contained within the Hill radii of fragment~1 and 2 are 
$\overline{M}_{\rm H,1}$=14.0~$M_{\rm J}$ and $\overline{M}_{\rm H,2}$=8.9~$M_{\rm J}$, 
respectively, indicating that if both fragments had survived
they would have formed massive GPs. 
The mean radii of fragments~1 and 2 are $\overline{R}_{\rm f,1}$=18~AU and 
$\overline{R}_{\rm f,2}$=15~AU, respectively, 
and their mean Hill radii are $\overline{R}_{\rm H,1}$=48.5~AU and 
$\overline{R}_{\rm H,1}$=41~AU. Both values are greater than the local scale hight $Z\approx25$~AU.

\begin{figure}
  \resizebox{\hsize}{!}{\includegraphics{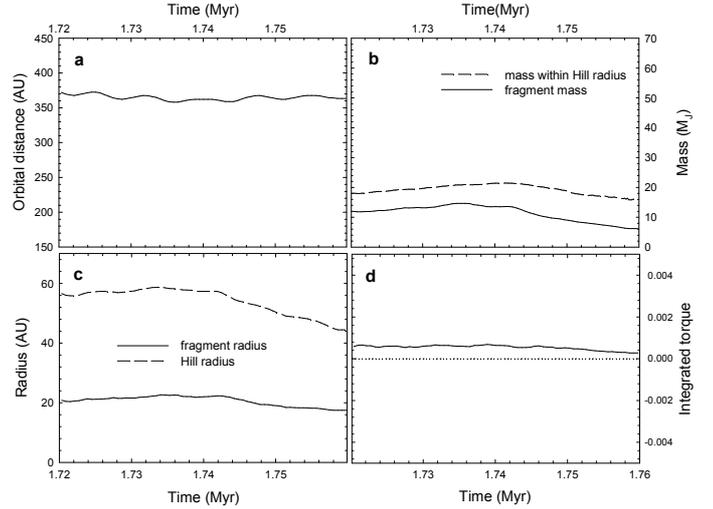}}
  \caption{Same as Figure~\ref{fig5} only for model~3.}
  \label{fig10}
\end{figure}

The likely reason why one of the fragments has dispersed is that this pair of
fragments violate the criterion for orbital stability between two coplanar planets on 
circular orbits \citep{Gladman93}
\begin{equation}
\triangle_{\rm f} \ge \triangle_{\rm f,cr} = 2\sqrt{3} R_{\rm H,M},
\label{stable}
\end{equation}
where $\triangle_{\rm f}=\overline{r}_{\rm f,2}-\overline{r}_{\rm f,1}=14$~AU 
and $R_{\rm H,M}$ is the mutual Hill radius defining the region in which gravitational 
force between two bodies is larger than the force on them
due to the star
\begin{equation}
R_{\rm H,M}= \left( {M_{\rm f,1} + M_{\rm f,2} \over 3 M_\ast} \right)^{1/3} 
{\overline{r}_{\rm f,1} + \overline{r}_{\rm f,2}
\over 2}.
\label{Mhill}
\end{equation}
Substituting the corresponding values into equations~(\ref{stable}) and (\ref{Mhill}), one obtains
$R _{\rm H,M}=56$~AU, $\triangle_{\rm f,cr}$=193~AU, and $\triangle_{\rm f} \ll \triangle_{\rm f,cr}$.
Evidently, the orbits of the two fragments are unstable. Moreover, the mutual Hill radius is
greater than the difference between the mean orbital distances of the fragments implying that 
the less massive fragment might not have withstood the disturbing tidal influence from the 
more massive counterpart. This effect might have been aided by insufficient numerical resolution
of our logarithmic polar grid at a radial distance of the fragments ($\sim 400$~AU). 
We note that the value of $\triangle_{\rm f,cr}=193$~AU is of the order of the orbital distance for
directly-imaged wide-orbit companions (see forth column in Table~\ref{table3}), which implies that
the separation between companions in multicomponent systems should be comparable to their orbital distances.
This makes the formation of such wide-orbit, multicomponent systems even more difficult. 
In any case, the results of this numerical simulation and other
studies \citep[e.g.][]{Boss11} have demonstrated that gravitational fragmentation can
account for the formation of multiple fragments at a time, but the question of whether these
fragments can ultimately mature into a system with more than one companion in wide orbits
is still open.

\begin{figure*}
 \centering
  \includegraphics[width=16cm]{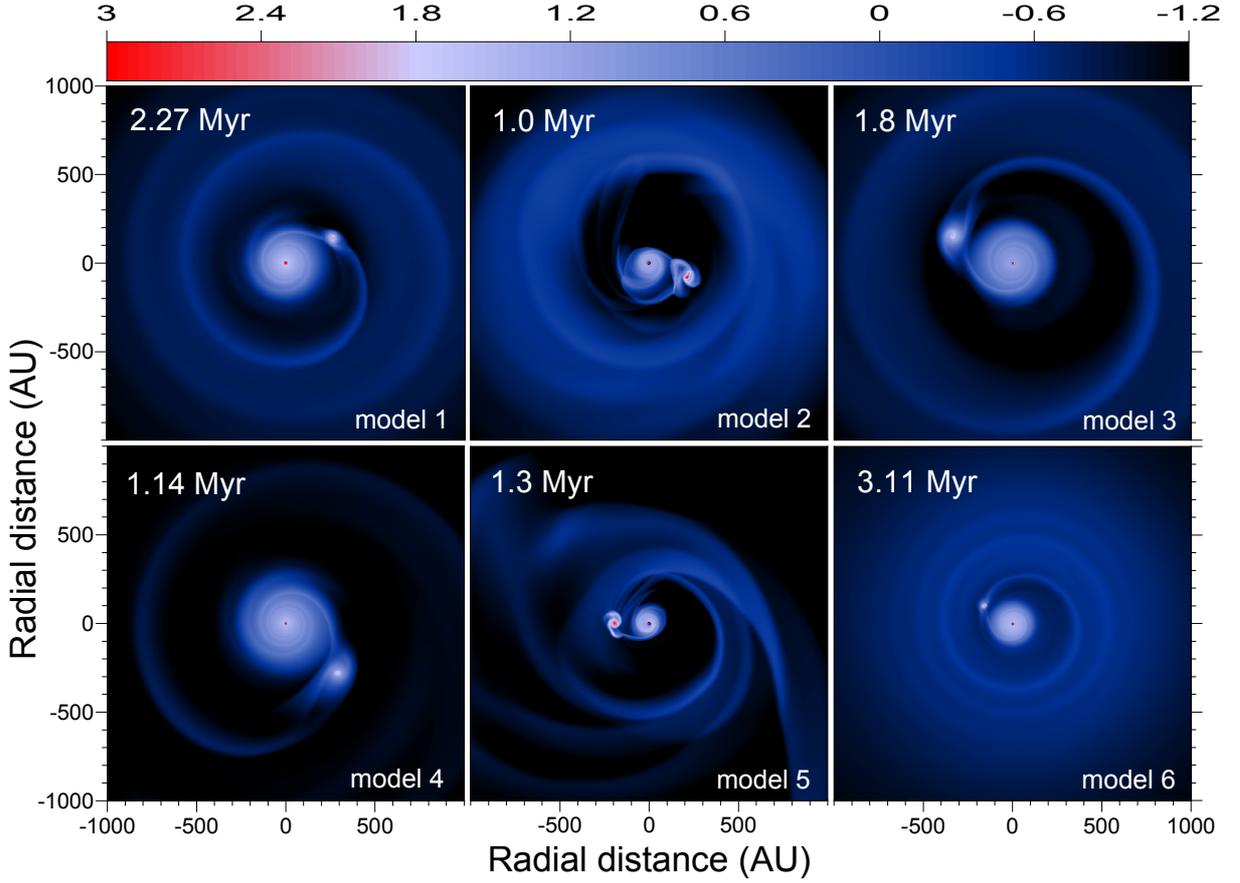}
  \caption{Gas surface density distribution in models showing the formation
  of quasi-stable GP/BD embryos on wide orbits. The model number and time elapsed
  since the formation of the central protostar is indicated in each panel. 
  Only the inner $2000\times2000$~AU box is shown. The scale bar is in log~g~cm$^{-2}$.}
  \label{fig11}
\end{figure*}

After the less massive fragment dispersed at $t\approx1.55$~Myr, the other 
fragment settled on a quasi-stable orbit with mean radial distance 
$\overline{r}_{\rm}=370$~AU. The main characteristics of the surviving fragment
are shown in Figure~\ref{fig10}, the layout of which is the same as that of
Figure~\ref{fig5}.
The last surviving fragment is characterized by the mean mass $\overline{M}_{\rm f}=11.0~M_{\rm J}$,
which is similar to the total mass of the two fragments before one of them dispersed. This
suggests that the surviving fragment has accreted most of the material released by 
the destroyed fragment.
Significant temporal variations in the instantaneous mass of the surviving fragment are indicative of
its highly perturbed state. The mean mass contained within the Hill radius 
is $\overline{M}_{\rm H}=19.0~M_\odot$.
The orbit of the surviving fragment is characterized by rather low
eccentricity, $\epsilon=0.02$. The integrated torque acting on the fragment is
positive and the estimated characteristic migration time of the fragment is $t_{\rm mg}=6.9$~Myr.
We conclude that this fragment is likely to evolve into a massive GP or low-mass BD,
depending on the amount of mass that will ultimately settle into a circumfragment disk.

\section{Characteristics of survived GP/BD embryos and comparison with observations}
\label{comparison}
We have run 60 models with the total integration time in each model 
exceeding 1.0~Myr after the formation of the central protostar. Protostellar disks
in the majority of  models were sufficiently massive to experience vigorous gravitational
fragmentation at radial distances greater than several tens of AU during the initial 
several hundred thousand years. The number of the fragments amounted to more than ten at a time.
However, most of the fragments have either migrated through the inner computational boundary
at 6~AU or 
got ejected from the computational domain into the intracluster medium or dispersed 
by tidal torques on time scales less than 1.0~Myr. 
Only six models out of 60 revealed the survival of one of the fragments
after 1.0~Myr of evolution.

\begin{table*}
\begin{center}
\caption{Characteristics of survived embryos}
\label{table2}
\renewcommand{\arraystretch}{1.5}
\begin{tabular}{c c c c c c c c c c c}
\hline \hline
model & $M_{\rm c}$ & $\beta$ & $M_\ast$ & $\overline{M}_{\rm f}$ & $\overline{M}_{\rm H}$ & 
$\overline{r}_{\rm f}$& $\epsilon$ & $\overline{R}_{\rm f}$ & 
$\overline{R}_{\rm H}$  & $t_{\rm mg}$ \\
\hspace{1cm} &  $(M_\odot)$ & (\%) & ($M_\odot$) &  ($M_{\rm J}$)   & ($M_{\rm J}$) & (AU) &  & (AU) & (AU) & (Myr) \\ [0.5ex]
\hline \\ 
1 & 1.7 & 0.56 & 1.2 & 11 & 20.5 & 330 & 0.07 & 20 & 47 & 10 \\
2 & 1.2 & 0.88 & 0.9 & 43 & 60 & 178 & 0.04 & 21 & 46 & 4 \\
3 & 1.5 & 0.56 & 1.1 & 11 & 19 & 370 & 0.02 & 20.5 & 54 & 6.9 \\
4 & 1.4 & 0.56 & 1.0 & 4.6 & 11.3 & 415 & 0.03 & 16.5 & 46 & 7 \\
5 & 1.55 & 1.27 & 0.75 & 27.5 & 40.5 & 180 & 0.06 & 19.5 & 41 & 3.9 \\
6 & 1.4 & 0.56 & 0.95 & 3.5 & 5.0 & 190 & 0.05 & 11.5 & 20.5 & 4.1 \\
\hline
\end{tabular}
\end{center}
\end{table*}

\begin{table*}
\begin{center}
\caption{Characteristics of known wide-orbit GPs and BDs}
\label{table3}
\renewcommand{\arraystretch}{1.5}
\begin{tabular}{c c c c c }
\hline \hline
Star & $M_\ast$ & $M_{\rm p}$ & $r_{\rm p}$& Age  \\
\hspace{1cm} & ($M_\odot$) &  ($M_{\rm J}$)   & (AU) & (Myr) \\ [0.5ex]
\hline \\ 
Oph 11 & 0.16 & $21\pm 3$ & $243\pm 55$ & $11\pm2$ \\
CHXR 73 & 0.35 & $15^{+8}_{-5}$ & 210 & 2 \\
DH Tau & $0.37 \pm 0.12$ & $11^{+3}_{-10}$ & 330 & 1 \\
CD-35 2722 & 0.4 & $31\pm 8$ & 67 & 100 \\
GSC 06214-00210 & 0.6 & $17\pm3$ & 320 & $11\pm2$ \\
Ross 458(AB) & 0.6 & $8.5\pm2.5$ & 1170 & $475\pm320$ \\
GQ Lup & 0.7 & $21.5 \pm 20.5$ & 103 & 1 \\
1RXS J1609 & 0.7 & $\approx 8$ & 330 & 2--5 \\
CT Cha & $0.8\pm0.1$ & 17 & 440 & $2\pm1$ \\
AB Pic & 0.8 & $13.5\pm 0.5$ & 260 & 30 \\
HN Peg & 1.0  & $16\pm 9$ & $795 \pm 15$ & 200 \\
HR 8799 & 1.6 & 5--10 & 15--68 & 20--150 \\ 
Fomalhaut & 2.1 & $3^{+1.2}_{-0.5}$ & 119 & 200 \\
\hline
\end{tabular}
\end{center}
\end{table*}

Figure~\ref{fig11} brings together six models that have demonstrated the formation of 
stable companions on wide orbits (to which we refer below as GP/BD embryos), 
showing for each model the gas surface density image (g~cm$^{-2}$ in log units) at 
the end of numerical simulations.
The model number and time elapsed since the formation of the central protostar
are indicated in each panel. Only the inner $2000\times2000$ box is shown for each model.
All six embryos possess their own circum-embryo disks, the masses of which are comparable to those 
of the parent embryos as implied by the mass contained within the Hill radius 
(see Table~\ref{table2} below). 
In particular, circum-embryo disks in models 2 and 5 exhibit a pronounces two-armed 
spiral structure. 

It is worth noting that the survived embryos and circum-embryo disks both exhibit 
a retrograde rotation with respect to that of the parental protostellar disk.
This goes against expectations of having prograde rotation \citep[e.g][]{Boley10,Zhu2012} 
borne out by the fact that the specific angular momentum of a fluid element 
in a rotationally supported disk increases with distance as $r^{0.5}$. 
Our numerical scheme performs well on standard numerical tests \citep{VB06}, including 
an angular momentum conservation test \citep{Norman80},
and hence this phenomenon is unlikely to be caused by numerical reasons.
However, significant local deviations from Keplerian orbits can
be present in strongly gravitationally unstable disks \citep[e.g.][]{Vor2010}, with a consequence 
that the specific angular momentum radial profile may not locally reflect a rotationally supported
disk. A visual analysis of Figures~\ref{fig2},
\ref{fig6}, and \ref{fig8} demonstrates that both the retrograde and prograde fragments
are present at the initial stages of disk evolution,
but the retrograde systems may be favoured for survival. For instance, the Type~III migration
depends strongly on the flow pattern near the planet \citep{Peplinski08}.
The counterrotating circum-embryo disk may make the formation of horseshoe streamlines near the corotation
more difficult, thus weakening the efficiency of Type~III migration.
In any case, a more focused study is needed to understand the phenomenon of retrograde rotation.

Embryos in models~1--5 satisfy the gap opening criterion~(\ref{gap}) and 
clear a well-defined gap in the protostellar disk. The embryo in model~6 is the least massive
of all and the corresponding gap is less pronounced.
We note that the gap is profoundly non-axisymmetric in models~2, 3, and 5, an effect that 
can in principle be used to infer the presence of massive GPs/BDs in the disk.

We present the main characteristics of surviving embryos in Table~\ref{table2}.
In particular, columns~1--11 list the model number, mass of the prestellar core
$M_{\rm c}$, ratio of rotational to gravitational energy in the prestellar 
core $\beta$, mass of the protostar $M_\ast$,
mean mass of the embryo $\overline{M}_{\rm f}$, mean mass within the Hill radius
$\overline{M}_{\rm H}$, mean orbital distance of the embryo $\overline{r}_{\rm f}$,
orbital eccentricity $\epsilon$, mean radius of the embryo $\overline{R}_{\rm f}$,
mean Hill radius $\overline{R}_{\rm H}$, and migration timescale $t_{\rm mg}$.
For the purpose of comparison, we also provide in Table~\ref{table3} 
the main characteristics of directly imaged, wide-orbit ($>50$~AU) companions 
to stars with mass $0.08~M_\odot \le M_\ast \le 2.1~M_\odot$. We excluded 
lower-separation companions because we are interested only in GPs/BDs 
that can be formed by disk gravitational fragmentation, which is likely to
occur at radial distances greater than 50~AU. In particular,
columns 2--5 list stellar masses $M_\ast$, masses of companions $M_{\rm p}$,
orbital distances of companions $r_{\rm p}$, and stellar ages as compiled 
by the Extrasolar Planets Encyclopedia (http://exoplanet.eu). We have ordered the
objects along the line of increasing stellar masses.

The embryo masses in our models lie in the $3.5-43~M_{\rm J}$ limits and cover
the entire range of masses found for detected wide-orbit companions. There is no clear 
indication that more massive cores tend to produce more
massive companions, though the range of companion-forming core masses considered
in the present study (1.2--1.7~$M_\odot$) 
is quite narrow. There is a hint that models with higher $\beta$ tend to produce more
massive companions (e.g., models 2 and 5) and that 
BD embryos tend to orbit less massive protostars,  but these tendencies need to
be confirmed with a wider sample of models. Our modelling suggests that 
the companion masses are likely to be determined by the disk/stellar properties
rather than by the properties of parental prestellar cores.
The masses contained within the Hill radius are generally a factor of 1.5--2.5 
greater than those of the embryos. The final masses of the embryos are therefore expected to
be somewhat higher, though the actual growth will be limited by the angular momentum of the
disk gas that the embryo is trying to accrete \citep{Boley10}.

The orbital distances of GP/BD embryos in our models are confined in the 178--415~AU limits,
whereas Table~\ref{table3} indicates a wider range of orbital distances for directly imaged companions,
15--1170~AU. While the upper limit on the orbital distance in our modeling (415~AU)
may increase if we run higher-resolution numerical simulations (due to
better resolution at large radial distances
on the adopted logarithmically-spaced radial grid), the lower limit (178~AU)
is not expected to change considerably given that all embryos in our models
are in fact slowly migrating outward. \citet{VB2010b} found GP embryos forming at distances
of the order 50~AU but their models employed a barotropic equation of state which is known
to facilitate disk fragmentation at small radial distances from the star.

The dearth of embryos
at orbital distances $\la170$~AU in our simulations cannot be explained by 
numerical resolution effects and is intriguing as it conflicts 
with observations. There are four objects in Table~\ref{fig3} that have companions
orbiting the star at radial distances less than 170~AU (CD-35 2722, GQ Lup, HR 8799, and Fomalhaut).
If disk fragmentation cannot explain these objects, then they must have been formed by 
dynamical scattering of closely-packed companions leading to ejection of the least
massive companion onto a wide orbit \citep[e.g.][]{Scharf09,Veras09}. 
The resulting orbits of these companions 
are expected to cover the full range of possible eccentricities, some of them may 
even become unbound with time \citep{Veras09}. In addition, the probability to form
wide-orbit companions via dynamical scattering declines quickly with the increasing
orbital distance of the companions \citep{Scharf09}.
The eccentricities of our formed embryos lie in the
0.02--0.07 range and the only known eccentricity of a wide-orbit companion is that
of Fomalhaut~b, $\epsilon=0.11$. Interestingly enough, the orbital distance of Fomalhaut~b is
119~AU and fits into the scattering scenario for the formation of wide-orbit 
companions. The above analysis suggests a unified picture for the formation of wide-orbit companions,
in which objects at orbital distances from several tens to $\approx 150$~AU are preferentially 
formed by dynamical scattering and are characterized by a whole spectrum of
eccentricities while those at greater orbital distances are
mostly formed via disk fragmentation and are characterized by low eccentricities.
Definite measurement of eccentricities for other wide-orbit companions
should therefore clarify the formation mechanism of GPs/BDs on wide orbits \citep[see also][]{DR09}.

The minimum mass of a protostar that hosts a wide-orbit GP/BD embryo in our
models is found to be $M_\ast=0.75~M_\odot$, whereas the directly-imaged 
wide-orbit companions have host stars with masses extending 
down to the brown-dwarf-mass regime \citep[e.g., UScoCTIO~108,][]{Bejar08}.
It appears that protostars with mass $\la 0.7~M_\odot$ possess
protostellar disks with mass that is insufficient to experience gravitational 
fragmentation in the T~Tauri phase. Fragmentation episodes in these disks are 
mostly confined to the embedded phase and are driven by mass infall 
from the parental core. As discussed in Section~\ref{fragmentation}, fragments in 
such disks have little chance to escape fast inward migration and are unlikely to
form wide-orbit companions. We note that \citet{Boss11} reported the formation
of companions in disks around stars with mass $<0.7~M_\odot$, but the total integration time
was limited to just 1000~yr and was insufficient to draw firm conclusions
about the survivability of these companions.

If disk fragmentation cannot explain wide-orbit companions around stars
with mass $\la 0.7~M_\odot$, then a viable alternative is dynamical
scattering of close-orbit companions.  
However, Table~\ref{table3} demonstrates that most of the 
stars with mass $<0.7~M_\odot$ 
have companions on orbital distances of the order of several hundred AU. 
Such large orbital distances
are difficult to explain in the framework of dynamical scattering since the
number of scattered objects quickly declines with increasing orbital distance \citep{Scharf09}.
This inconsistency necessitates further research into this subject.

Finally, we note that our modelling failed to produce systems with 
more than one stable companion. This is consistent 
with observations. The only known system with several wide-orbit companions is
HR~8799 and even in this case only one companion has an orbital distance
greater than 50~AU. The lack of multiple wide-orbit companions
is likely caused by the fact that the orbital stability criterion~(\ref{stable}) 
imposes strict limitations on the minimum orbital distance between
two stable companions $\triangle_{\rm f, cr}$. 
For instance, in model~3 that showed an attempted formation
of two companions (but one of them finally dispersed), 
the corresponding minimum distance was $\triangle_{\rm f,cr}=193$~AU,
while the actual mean radial separation between the companions
was only 14~AU. Large values of $\triangle_{\rm f,cr}$ 
make the disk fragmentation scenario problematic for 
explaining multi-component systems
as it may require very extended and hence massive disks.

\section{Conclusions}
\label{summary}
We computed the gravitational collapse of prestellar cores with 
masses lying in the $0.1~M_\odot<M_{\rm c}\le1.8~M_\odot$ range and ratios of rotational 
to gravitational energy confined in the $0.2\% < \beta \le 2.2\%$ limits. The 
integration time in our numerical hydrodynamics simulations
extended beyond 1~Myr after the formation of the central protostar and covered
the entire embedded phase and part of the T~Tauri phase of stellar evolution.
We focused on models that showed disk gravitational fragmentation and, in 
particular, on models that revealed the formation of quasi-stable, giant planet (GP) 
and brown dwarf (BD) embryos on orbits greater than 50~AU 
(referred as wide-orbit companions). The typical migration timescales of the embryos
are comparable to or greater than the lifetime of a typical disk (2--3~Myr),
which allows us to conclude that they will ultimately
cool and contract into fully formed GPs or BDs.
Insufficient numerical resolution does not allow us
to resolve the formation of planetary-sized objects.

While most of our models showed disk fragmentation, only
6 out of 60 models revealed the formation of wide-orbit companions.
We compared the characteristics of our embryos with those of fully formed GPs
and BDs obtained from direct imaging (http://exoplanet.eu). 
The disk masses 
and radii provided below were calculated by time-averaging instantaneous 
values over the duration of the Class I phase, which is most relevant to studying 
disk fragmentation. Our findings can be summarized as follows.
\begin{itemize}

\item Masses of wide-orbit companions lie in the $3.5-43~M_{\rm J}$ limits 
and cover  the entire range of masses found for directly-imaged GPs and BDs
on orbits greater than 50~AU. There is no clear indication that the mass of the companion
depends on the mass and angular momentum in the pre-stellar core, 
though a wider sample of companion-forming
models is needed to draw firm conclusions.

\item The orbital distances of the companions found in our modelling
lie in the 178--415~AU limits. This range of orbital distances 
is notably narrower than that found for directly-imaged companions,
a few~AU--1170~AU. While the upper found limit (415~AU) may increase if we
consider higher resolution simulations\footnote{Numerical resolution
deteriorates at large distances on our logarithmically
spaced radial grid.}, the lower limit (178~AU) cannot be explained by resolution effects
and is intriguing as it conflicts the observations. We propose that
companions at orbital separations from a few tens to 150~AU are 
likely to form via dynamical scattering of closely-packed companions
onto wide orbits, while companions at larger orbital distances are predominantly
formed via disk fragmentation. Definite measurement of eccentricities 
as a function of orbital distance
should clarify the formation mechanism of GPs/BDs on wide orbits,
because the disk fragmentation scenario tends to produce companions
on low eccentricities, $\epsilon\le 0.1$.

\item Our numerical simulations did {\it not} produce multiple companions on wide orbits. 
Although model~3 revealed an attempted formation of two companions
at orbital distances $\approx 190-210$~AU, one of the fragments 
dispersed in less than 0.5~Myr. The likely reason for this failure 
is that the radial separation between the two companions, $\approx 14$~AU, 
was much smaller than the required 190~AU according to 
the  orbital stability criterion of \citet{Gladman93}. 
The formation of multi-component systems 
would require very extended and hence massive disks, which are statistically rare.
The only known system with several wide-orbit companions, HR~8799, has
only one companion at a distance greater than 50~AU and is likely to 
form via dynamical scattering.

\item The minimum mass of a companion-hosting star found in our 
modelling is 0.75~$M_\odot$, whereas the corresponding value
for directly imaged systems extends down to the BD-mass regime. 
It is likely that disk fragmentation in systems
with stellar mass $\le 0.7~M_\odot$ is mostly confined to the embedded
phase of star formation, in which mechanisms causing 
migration/ejection/destruction of the fragments are strong
and the likelihood for survival of the fragments is low.

\item Disk gravitational fragmentation does {\it not} automatically 
guarantee the formation of wide-orbit companions. Most of the fragments
do not stay on wide orbits for more than several orbital periods. 
The majority are torqued into the disk inner region and through the 
sink cell (6~AU) by gravitational torques  
from spiral arms or other fragments, a few fragments may be ejected from 
the disk via many-body gravitational interaction
\citep{Stamatellos09,BV12}, and some
fragments may be tidally dispersed at various radial distances in the disk 
\citep{Boley10,Nayakshin10,Vor2011a,Zhu2012}. 
The fragments that pass through the sink cell 
may be completely destroyed and accreted onto the forming
star \citep{VB06,VB2010a,Machida2011} or lose their gaseous envelopes and 
form terrestrial cores or icy giants if
the dust sedimentation timescale was sufficiently short \citep{Nayakshin10,Boley10}.

\item The minimum disk mass at which fragmentation can take place 
is found to be $\overline{M}_{\rm d}^{\rm fr}=0.07~M_\odot$ for models 
with $\beta\ga 1.2\%$. 
The value of $\overline{M}_{\rm d}^{\rm fr}$ may increase by almost 
a factor of 2 for disks formed from pre-stellar cores with lower angular momentum, 
$0.2\% \le \beta < 1.2\%$. Our found values of $\overline{M}_{\rm d}^{\rm fr}$ 
are in reasonable agreement with previous estimates of the critical disk mass, 
$\approx 0.1~M_\odot$ \citep{Rice03,Mayer07}.

\item The minimum disk mass that is required for the formation of wide-orbit companions
is found to be $\overline{M}_{\rm d}^{\rm c.f.}=0.21~M_\odot$,
a factor of 3 greater than the minimum disk mass for 
gravitational fragmentation $\overline{M}_{\rm d}^{\rm f}=0.07~M_\odot$.
The minimum disk radius that is required to produce wide-orbit 
companions is 370~AU, more than a factor of two greater than that required for
disk fragmentation to occur.
\end{itemize}

The overall low probability of the fragment survival and the efficient migration/destruction
mechanisms found in the present study both support the recently proposed disruption/downsizing
hypothesis \citep{Boley10,Nayakshin10}. Our results are also in agreement with
the recent study of \citet{Zhu2012}, who studied numerically gravitational fragmentation
in two-dimensional disks and found that only 3 out of 13 fragments escaped fast
inward migration. 
In the present paper, we have ignored a possible motion of the central star due to 
a combined gravitational potential of the non-axisymmtric disk. Analytic 
studies suggest that the stellar wobbling may increase the strength of the odd spiral 
modes, and in particular that of the $m = 1$ mode \citep[e.g.][]{Shu90}, which 
may increase the gravitational torque exerted on the fragments and make it even more difficult
for the fragments to survive. On the other hand, focused numerical hydrodynamic simulations by \citet{Michael10}
found no evidence for the enhanced $m=1$ mode in a $0.14~M_\odot$ disk around a solar-mass star.
The effect of stellar wobbling in more massive disks, relevant for the present results, needs to be
considered in a future study.

 \section{Acknowledgments}
The author is thankful to the referee for useful comments which helped to improve the paper.
The author also thanks Prof. Shantanu Basu for stimulating discussions.
This work is supported by the RFBR grants 10-02-00278 and 11-02-92601. The simulations were performed
on the Shared Hierarchical Academic Research Computing Network (SHARCNET),
on the Atlantic Computational Excellence Network (ACEnet), and on the Vienna Scientific
Cluster (VSC-2).

\end{document}